\documentclass[10pt, twocolumn]{article}

\usepackage{lipsum}

\usepackage{styles/style}
\usepackage{styles/QIA-network-diagram-style-guide}

\title{A Modular Quantum Network Architecture for Integrating Network Scheduling with Local Program Execution.}

\author[1,2,3]{Thomas R. Beauchamp}
\author[1,2,3]{Hana Jirovská}
\author[1,2,3]{Scarlett Gauthier}
\author[1,2,3]{Stephanie Wehner}

\affil[1]{QuTech, Delft University of Technology}
\affil[2]{Kavli Institute of Nanoscience, Delft University of Technology}
\affil[3]{Quantum Computer Science, Electrical Engineering, Mathematics and Computer Science, Delft University of Technology}

\date{}

\makeatletter
\addtotheorempostheadhook[define]{\renewcommand{\thmt@thmname}{\protect\refstepcounter{define}}}
\addtotheorempostheadhook[nsmodel]{\renewcommand{\thmt@thmname}{\protect\refstepcounter{nsmodel}}}
\makeatother

\begin{document}

\maketitle

\begin{abstract}
We propose an architecture for scheduling network operations enabling the end-to-end generation of entanglement according to user demand. The main challenge solved by this architecture is to allow for the integration of a network schedule with the execution of quantum programs running on processing end nodes in order to realise quantum network applications. A key element of this architecture is the definition of an entanglement packet to meet application requirements on near-term quantum networks where the lifetimes of the qubits stored at the end nodes are limited. Our architecture is fully modular and hardware agnostic, and defines a framework for further research on specific components that can now be developed independently of each other. In order to evaluate our architecture, we realise a proof of concept implementation on a simulated 6-node network in a star topology. We show our architecture facilitates the execution of quantum network applications, and that robust admission control is required to maintain quality of service.
Finally, we comment on potential bottlenecks in our architecture and provide suggestions for future improvements.
\normalfont
\end{abstract} 
\section{Introduction}\label{sec:introduction}

The realisation of a quantum internet will enable the use of new networked applications beyond what is possible with the current classical internet. 
Such applications include the ability to perform verifiably secure secret sharing \cite{ekert_quantum_1991, BB84}, secure remote computation \cite{bqc1, bqc2, childs_secure_2005} and securely electing a leader \cite{tani_exact_2012}, amongst many others \cite{wehner_quantum_2018}. 
The aim of any quantum network architecture therefore should be to ensure that these applications can be successfully executed. 

To execute a quantum application, so-called \textit{entangled links} between the \textit{end nodes}, the quantum devices the users have access to, are required. 
Each of these entangled links is a pair of entangled qubits with a fidelity with respect to an  EPR pair \cite{einstein_can_1935}, where the fidelity is a measure of the quality of the entangled link \cite{nielsen_quantum_2010}. 
However, such links are difficult to produce and doing so requires the use of a limited quantity of resources in the quantum network \cite{dur_quantum_1999, pompili_realization_2021, inesta_optimal_2023, theoryHeraldCCGZ}. 
Furthermore, at present each link has a limited usable lifetime as quantum memories experience decoherence over time, reducing the quality of stored entangled links \cite{bradley_robust_2021}. 
Therefore, there exist two interacting scheduling problems which must be solved: firstly how should the limited network resources be assigned to pairs of users to allow them to generate entangled links (the \textit{network scheduling problem}), and secondly how to efficiently schedule the execution of quantum applications to efficiently use any entangled links which are generated (the \textit{local scheduling problem}). 

The first operating system for end nodes, QNodeOS \cite{QNodeOS}, allows the second of these problems to be solved. \textit{Qoala} \cite{vecht_qoala_2025}, an application execution environment which runs on QNodeOS, offers an improved end node execution environment.
QNodeOS breaks an application into a program at each node, which is then further subdivided into blocks of instructions that can be scheduled for execution at runtime. 
Furthermore, Qoala enables a compiler which can provide advice about the quantity and quality of entangled links which are required. 
Although this application execution environment successfully address the local scheduling problem, they require the existence of a network schedule to supply allocations of time during which entanglement generation can take place. 

However, there presently does not exist a network architecture which can produce schedules which are compatible with such an execution environment.
Without such an architecture, the network scheduling problem cannot be solved in a manner which still allows the local scheduling problem to be solved effectively. 
In this work we therefore propose such an architecture to unify the approach to the network and local scheduling problems. In particular:

\begin{itemize}
    \item \textbf{We introduce the first quantum network architecture which takes a unified approach to scheduling the execution of quantum applications on end nodes and scheduling the use of resources on the network}. 
    As part of this we introduce a notion of \textbf{packets of entanglement} to capture the requirements on entangled links imposed by applications, and a corresponding notion of a \textbf{packet generation task} to allow the network to efficiently schedule time for the generation of these packets of entanglement. 
    We also define a \textbf{demand format} which captures all the information required by the central controller to be able to compute network schedules.
    This architecture also provides a \textbf{modular framework} within which further research can be undertaken to develop synergistic network and local scheduling strategies. 
    This will enable network schedules to be integrated into local program execution in a well-defined and consistent manner.
    \item \textbf{We provide an example implementation of our architecture in simulation}, using earliest deadline first derived methods for network scheduling. We use this implementation to perform numerical simulations of our architecture and create a baseline performance evaluation against which to benchmark further work in the domain of quantum network scheduling. 
      The code used for this implementation and the data used in the evaluation is available from \cite{beauchamp_data_2025}.
\end{itemize}

\subsection{Structure}
The rest of this paper is structured as follows: In \ref{sec: Related Work} we give some background and related work and discuss how our work here differs.
In~\ref{sec: Design Considerations} we discuss some of the design considerations which inform the design of our architecture.
In~\ref{sec: Network Architecture} we lay out our proposed network architecture.
In~\ref{sec: Example Implementation} we give an example of an implementation of the architecture.
Finally in \ref{sec: Evaluation} we evaluate the performance of our architecture using a specific implementation, and in \ref{sec: conclusions and future work} we comment on future directions for research. 
A table summarising all the notation we use in this paper can be found in Appendix~\ref{app: notation}.

\section{Background}
\label{sec: Related Work}
\subsection{Network Model}

There are many different models for how a quantum network should operate. 
In the literature, one common example is what we will call a \textit{pre-loaded} network, where there is a high probability that entangled links are immediately available to an application (e.g.  \cite{ESDI, pirker_quantum_2019}). 
Such networks rely on continuously generating and buffering entanglement between each pair of network components. 
However, restrictions on achievable entanglement generation rates, buffer lifetimes and buffer capacities limit the possibility of near-term implementations of such networks.
For example, experiments on leading hardware \cite{pompili_experimental_2022} have realised a three-node network, on which they reported memory lifetimes of 11ms and generation rates of one end-to-end link every 40s. 
Therefore, we instead consider a \textit{generate-when-requested} network. 
In such networks, we do not assume that end-to-end entangled links can be stored between any two scheduled periods of time.
This type of network is implementable with the technological maturity of current devices and those that will exist in the coming years.

\subsection{Related Work}

Our architecture is designed to be compatible with the network application execution environment Qoala, which is described in~\cite{vecht_qoala_2025}.
Qoala is an extension to the QNodeOS operating system for quantum network end nodes developed by Delle Donne \textit{et al.} in \cite{QNodeOS}. 
When using the Qoala execution environment on QNodeOS, it is assumed that there exists a network schedule in order for entanglement generation to occur. 
One of the aims of this work is to design an architecture which can produce suitable  network schedules which enable the execution of quantum network applications using Qoala on QNodeOS. 

%

Our work is also compatible with the network stack proposed in Dahlberg \textit{et al.} in \cite{dahlberg_link_2019}.
In particular, the network schedules that our architecture produces replace the queue used in the implementation of \cite{dahlberg_link_2019} in \cite{pompili_experimental_2022}.

We build on the formalism presented in \cite{davies_tools_2024}.
In particular we incorporate the framing of entanglement generation as a problem in the field of scan statistics (see e.g. \cite{naus_approximations_1982, glaz_scan_2001}) into our architecture.

There are several other authors who have proposed architectures for quantum networks, however they each have a different scope for the architecture than ourselves.
In particular, none of them explicitly consider the influence of requirements of executing local applications on the network.

For instance, Skrzypczyk \textit{et al.} in \cite{skrzypczyk_architecture_2021} propose an architecture around using TDMA schedules to generate good quality entanglement.
Whilst we build on their ideas about scheduling, they do not consider how their scheduling will impact the ability of end nodes in the network to execute the applications requiring this entanglement, nor how the demands are generated from the applications.

Cicconetti \textit{et al.} in \cite{cicconetti_request_2021} and Gu \textit{et al.} in \cite{ESDI} consider the problem of scheduling requests for entanglement generation in a quantum network.
However in both cases they consider a continuous distribution network rather than a generate-when-requested network.
Furthermore, they do not consider how the end nodes use the entanglement which is generated, and also consider a system where every edge in the network is in constant communication with the central controller, whereas we only require sporadic communications.
A disadvantage of this compared to our approach is that due to latency when communicating between the edges and the central controller, there is an inherent loss in the quality of entangled links which can be created. 
Furthermore, they only consider requests for single entangled links, whereas we consider requests for packets of entangled links.

Van Meter \textit{et al.} in \cite{van_meter_quantum_2022} propose an architecture which focuses on routing of entanglement across many smaller networks, and the protocols required to do so.
This again does not account for the local nodes, nor does it consider the interaction of scheduling entanglement generation on the network with executing the applications on end nodes.
Furthermore, the scope of our work is to provide an architecture for a single quantum network which can be centrally controlled, rather than for a quantum internetwork.  

Our architecture also has similarities to a software-defined network \cite{kreutz_software-defined_2015, mckeown_openflow_2008, halpern_forwarding_2010}. 
In particular, we see our work as a method of facilitating the implementation of a quantum SDN. 
There have been several examples of prior work on defining a quantum SDN.
For example in \cite{kozlowski_p4_2020}, Kozlowski, Kuipers and Wehner give an implementation of a quantum SDN using the P4 language. 
However, this implementation only focuses on the network aspects, and does not consider the execution of applications on the end nodes. 
There has also been a quantum SDN proposed and demonstrated by Yang and Cui in~\cite{yang_reconfigurable_2023}.
However again this work does not explicitly consider scheduling at the end nodes in the network, and they focus on demands being registered via a web-interface.
In contrast we create a fully-autonomous architecture, where the end nodes, rather than the users themselves, submit the demands in response to users wishing to run specific applications.

There have also been several SDN architectures proposed for quantum key distribution (QKD), e.g.~\cite{tessinari_software-defined_2023,martin_madqci_2023,aguado_quantum-aware_2016,yu_software_2017,aguado_secure_2017,wang_quantum-key-distribution_2019,wang_network_2023,hadi_quantum_2024,iqbal_sdn-enabled_2024,gupta_chaqra_2024}. However, our architecture can support arbitrary applications, rather than just QKD.
This imposes more constraints on the process of generating entanglement than are considered in these works.
For the same reason we cannot use the demand format in \cite{quantum_key_distribution__etsi_industry_specification_group_quantum_2024}.
Furthermore, we are focused on the execution of these applications rather than adding extra security to a classical network as in \cite{tessinari_software-defined_2023}.

Our architecture also has similarities to a so-called \textit{Time-triggered Ethernet (TTEthernet)} architecture~\cite{craciunas_combined_2016}.
However, for our problem one cannot just use a classical TTEthernet system for a couple of reasons.
Firstly, as a classical system the literature does not directly take into account quantum-specific constraints such as decoherence.
Secondly, the predominant usages of TTEthernet in the classical sphere are in control systems,  such as in spacecraft~\cite{kramer_ttethernet_2022,tttech_nasas_2023}. where the sources of (time triggered) demands on the network are known \textit{a priori} and can therefore be accounted for.
However, in our model the network does not know where demands will come from, what resources or objectives they will require, or for how long they will need to use the network.

\section{Design Considerations}
\label{sec: Design Considerations}

Our network architecture defines a framework for integration of a network schedule with the execution of quantum programs running on end nodes. The design of the architecture thus inherits considerations pertaining to robust operation of a quantum network (see e.g. \cite{dahlberg_link_2019}) and considerations from the application execution environment of end nodes \cite{vecht_qoala_2025}. We describe at a high level each of the relevant principles and highlight how they may be consistently combined into the foundational pillars of a single network architecture.

\begin{figure*}
    \centering
    \resizebox{!}{!}{

\begin{tikzpicture}[
            node distance = 1cm,
]

            \node[processingNode] (A) {$B$};
            \node[metropolitanHub] (B) [right = 1cm of A] {$M_1$};
            \node[PN] (X) [above of = B, xshift = -1cm] {$A$};
            \node[processingNode] (Y) [below of = B,xshift = -1cm] {$C$};

            \node[jct] (J1) [right=0.2cm of B] {$J_1$};
            \node[jct] (J) [right=2cm of J1] { $J_2$};
            \node[jct] (J2) [above=0.4cm of J] {$J_3$};
            \node[jct] (J3) [right=4cm of J] {$J_4$};

\node[metropolitanHub] (M2) [right = 0.2cm of J3] {$M_2$};
            \node[processingNode] (E) [below of = M2,xshift = 1cm] {$D$};
            \node[processingNode] (U) [above of = M2,xshift = 1cm] {$E$};

\node[PN] (T) [above=0.5cm of J2] {$F$};

            \path let \p1 = ($(J)!0.5!(J3)$) in coordinate (midpoint M2 M3) at (\p1);

            \node[RN] (3) [below = 0.7cm of midpoint M2 M3] {};
            \node[RN] (2) [left=0.2 cm of 3] {};
            \node[RN] (1) [left = 0.2cm of 2] {};
            \node[RN] (4) [right= 0.2cm of 3] {};
            \node[RN] (5) [right=0.2 cm of 4] {};

            \node[] (0) [left = 0.4cm of 1] {};
            \node[] (7) [right = 0.4cm of 5] {};

            \draw (A.east) -- (B.west);
            \draw (X.south east) -- (B.north west);
            \draw (Y.north east) -- (B.south west);
            \draw (B.east) -- (J1.west);
            \draw[decorate, decoration=zigzag] (J1.east) -- (J.west);
            \draw[decorate,decoration=zigzag] (J.east) -- (J3.west);
            \draw (J3.east) -- (M2.west);
            \draw[decorate,decoration=zigzag] (J.north) -- (J2.south);
            \draw[decorate,decoration=zigzag] (J.south) -- ([yshift=-1cm]J.south) node [below, yshift=0.2cm] {$\vdots$};
\draw (M2.south east) -- (E.north west);
            \draw (M2.north east) -- (U.south west);

            \draw (J2.north) -- (T.south);

            \draw (0.east) -- (1.west);
            \draw (1.east) -- (2.west);
            \draw (2.east) -- (3.west);
            \draw (3.east) -- (4.west);
            \draw (4.east) -- (5.west);
            \draw (5.east) -- (7.west);

            \draw[decorate, decoration = {brace,mirror,raise = 5pt, amplitude = 6pt}] ([xshift=0.1cm]J.east) -- ([xshift=-0.1cm]J3.west);
            \draw[decorate, decoration = {brace,raise = -18pt, amplitude = 6pt}] ([xshift=0.1cm]J.east) -- ([xshift=-0.1cm]J3.west);

\end{tikzpicture}

 }
    \caption[Example of a Quantum Network.]{A general quantum network may be built from four kinds of devices: end nodes, metropolitan hubs, repeater chains, and junction nodes (see \ref{subsubsec: DC - Network - Devices and Components}). These devices may be connected in the manner illustrated. Circles $A$-$G$ represent user controlled end nodes, squares $M_i$ are metropolitan hubs and diamonds $J_i$ are junction nodes. Repeater chains are represented by zig-zag lines, while individual quantum repeater nodes are represented by triangles. }
    \label{fig:Example Quantum Network}
\end{figure*}
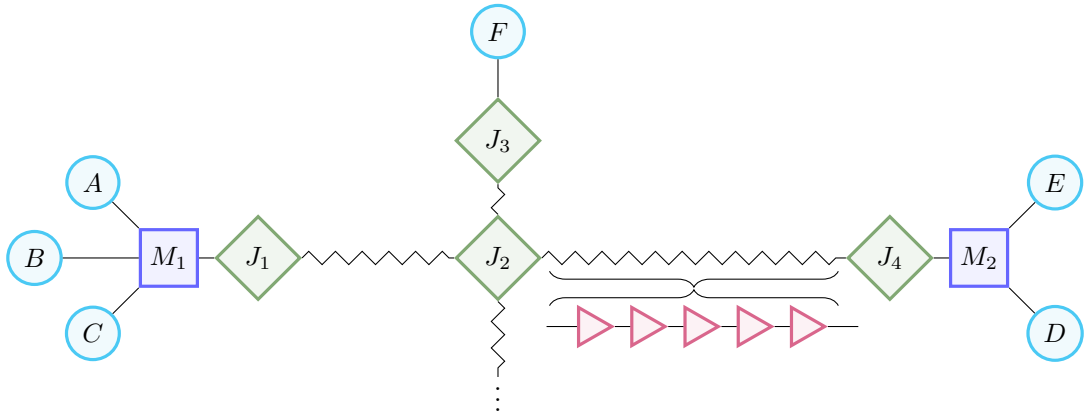

\subsection{Network}

\subsubsection{Devices and Components}
\label{subsubsec: DC - Network - Devices and Components}
We consider a quantum network comprised of four types of devices. Figure~\ref{fig:Example Quantum Network} illustrates an example of how these components can fit together.
The first device type, called \textit{end nodes}, execute quantum applications, operate under independent (local) control, and accept input from users. 
An end node may be a processing node with some memory capabilities, such as in \cite{ruf_quantum_2021, ThreeNodeQN, HerEntTrappedIons2, HerEntTrappedIons1, NeutralAtoms1, NeutralAtoms2, HerEntOriginalAE, HerEntSecondAE}, or a device which is capable of preparing/measuring single photons, such as in \cite{PhotonicClients1, MDIQKDDeployed}. Any end node can also perform classical operations, such as arithmetic operations and classical communication. 

End nodes may be connected to a second type of device, \textit{metropolitan hubs}. 
These are devices which enable pairs of nodes located close together (typically <50km end-to-end) to create entangled links.
Examples of such devices are entanglement distribution switches \cite{EDS1GVstochastic, EDS2GVexact, EDS3}, which employ quantum memories at the hub, or alternatively entanglement generation switches \cite{gauthier_control_2023, EGS2}, which do not rely on quantum memories at the hub.
A metropolitan hub will typically have many devices connected to it, including both end nodes and repeater chains. 

The third type of component  we include are \textit{repeater chains}, which allow for long distance entangled links to be created between two parties. 
Repeater chains are made from a linear chain of \textit{repeater nodes}, such as those in \cite{theoryHeraldCCGZ, theoryHeraldDLCZ, RepeatersPhotonic1, RepeatersAE1, RepeatersTrappedIon1}.
We treat repeater chains as a single network component which can be configured to produce links between two border nodes connected to either end of the repeater chain at a fixed rate and average fidelity.

The final component we consider in the network are \textit{junction nodes}. 
These provide an interface between multiple repeater chains and between repeater chains and a metropolitan hub. 
In the latter case, junction nodes are also referred to as border nodes \cite{maiti_border_2024}.
Such nodes remove the requirement to have direct repeater chains between every pair of metropolitan hubs. Junction nodes may be implemented using a combination of one or more of the previous devices.

We refer to repeater chains,  metropolitan hubs and junction nodes collectively as \textit{internal} components, and end nodes as \textit{external} components. 

We assume that each internal component of the network has a control API which can be used to install network schedules.
Furthermore, such an API is able to expose information about the operational parameters of the component. 
For example, a metropolitan hub may expose the maximum number of pairs of node it is possible to simultaneously connect; a repeater chain may expose the rate and fidelity at which entangled links between the end points are created; and junction nodes may expose the number of links which can be stored and for how long. 

\subsubsection{Hardware Agnosticism}
Each of the multiple possible implementations of any network device is associated with specific requirements relating to its operation. However, a network architecture should be able to seamlessly support whatever hardware is being used. As a result, where necessary we assume that the control API of internal network components and the application execution environment of end nodes make concessions for the specific requirements of a device, ensuring correct functioning. Thus our network architecture is hardware agnostic and compatible with a heterogeneous network consisting of devices based on a variety of different hardware types.

\subsubsection{SDN Controller}

 \textit{Software defined networks} (SDN)s \cite{kreutz_software-defined_2015,halpern_forwarding_2010, mckeown_openflow_2008} separate the data plane of a network, which forwards traffic to the appropriate destinations, from the control plane, which makes decisions about how traffic should be handled. 
 Decisions taken by the control plane include routing and resource access management.
 The SDN framework aims to simplify network management and make networks flexible and cost-effective. 
 We consider the traffic on the data plane of a quantum network to consist of point to point attempts to generate entanglement.
 
We assume that the quantum network follows the general architecture of an SDN.
In particular, we assume that the network is overseen by a (logically) central controller, similar to for example \cite{martin_madqci_2023, kozlowski_p4_2020, skrzypczyk_architecture_2021, pompili_realization_2021}.
Such a controller has the authority to compute and enforce \textit{network schedules}, which dictate when entangled links can be generated for particular pairs of end nodes. 

We also assume that the central controller has a complete overview of the entanglement generation capabilities of the components of the network (consideration \ref{design considerations: Network capabilities}).
However, we do not assume it knows whether or not any given attempt to generate an entangled link succeeds or fails.

\subsubsection{Device Autonomy}

We assume that each component of the network is able to operate without direct interaction with the central controller.
In particular, we assume for each component in the network there is a local controller which handles executing an installed network schedule on that component without further input from the central controller.
On end nodes, this local controller takes the form of an execution environment such as Qoala \cite{vecht_qoala_2025} running on QNodeOS \cite{QNodeOS}.
On internal components, this local controller may be more limited and simply implement a rule-set which realises the network schedule. 
For example, if a repeater chain is required by the network schedule to produce entangled links with certain rate/fidelity characteristics, then we assume there exists a logically centralised local controller which implements some policy, e.g. \cite{inesta_optimal_2023, kamin_exact_2022}, on the repeater chain in order to create the required entangled links.

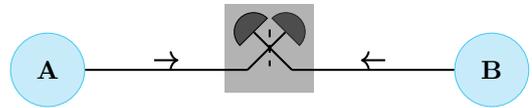
\begin{figure}
    \centering
    \resizebox{0.8\linewidth}{!}{
\begin{tikzpicture}
    \node[rectangle] [minimum width=1.2cm, minimum height = 1.2cm, fill = black!30] at (0,0) {};
    \node[semicircle, minimum width = 0.5cm, rotate = 45, fill = black!70, draw=black] (h1) at (-0.3, 0.3) {};
    \node[semicircle, minimum width = 0.5cm, rotate = -45, fill = black!70, draw=Black] (h2) at (0.3, 0.3) {};

    \draw[thick, dashed] (0, 0.25) -- (0, -0.25);

    \draw[thick] (h1.south) -- (0.3, -0.3);
    \draw[thick] (h2.south) -- (-0.3, -0.3);

    \node[circle, minimum width=1cm, draw=ProcessBlue!60, fill = ProcessBlue!20] (e1) at (-3, -0.3) {$\mathbf A$};
    \node[circle, minimum width=1cm, draw=ProcessBlue!60, fill = ProcessBlue!20] (e2) at (3, -0.3) {$\mathbf B$};

    \draw[thick] (e1.east) -- (-0.3, -0.3) node [above, midway, name=e1arrow, minimum width = 0.3cm] {};
    \draw[thick] (e2.west) -- (0.3, -0.3) node [above, midway, name=e2arrow, minimum width = 0.3cm] {};

    \draw[->, thick] (e2arrow.east) -- (e2arrow.west);
    \draw[->, thick] (e1arrow.west) -- (e1arrow.east);

\end{tikzpicture} }
    \caption[Example of an entanglement generation protocol.]{Example of an entanglement generation protocol (heralded entanglement). Nodes $A$ and $B$ probabilistically send photons to a `heralding station' located midway between them. This heralding station consists of (single) photon detectors and a beamsplitter. Depending on the pattern of photons detected, it is possible to determine if an elementary entangled link has been generated, and to subsequently inform the nodes. More details can be found in, e.g. \cite{theoryHeraldCCGZ}.}
    \label{fig:single click diagram}
\end{figure}

\subsubsection{Protocols for generating entanglement}

Our architecture is agnostic to how entangled links are produced at the physical layer. 
However, we assume that any protocol which is used allows  the rate and fidelity at which entangled links can be produced to be calculated and exposed. 
We also assume that entangled links between neighbouring nodes are created using a heralded entanglement generation scheme such as \cite{theoryHeraldCCGZ, theoryHeraldDLCZ} (Figure \ref{fig:single click diagram}).
In particular this means that the nodes attempting to create an entangled link receive a success or failure outcome, indicating when an entangled link has been created. Hence subsequent attempts may be triggered conditioned on the success or failure of previous attempts.

We refer to entangled links between end-nodes as \textit{end-to-end entangled links}, and entangled links between neighbouring nodes as \textit{elementary entangled links}.

Note that in the literature an \textit{entanglement generation attempt} typically refers to an attempt to generate an elementary entangled link, following some specified entanglement generation protocol, for example as in Figure \ref{fig:single click diagram}.
However, when referring to entanglement generation attempts we will mean attempts to generate end-to-end entangled links, requiring the use of all internal components along a specified path and employing a pre-determined protocol. 
For example, attempts to generate entanglement between nodes $A$ and $D$ in Figure~\ref{fig:Example Quantum Network}, require the simultaneous use of resources at $M_1$, $J_1$, $J_2$, $J_4$ and $M_2$, as well as the repeater chains connecting these components.

\subsubsection{Network Stack}
 A network stack is a layered set of protocols and services that work together to enable network communication. Each layer in the stack is responsible for a specific aspect of network functionality. Together the layers provide a comprehensive framework for data transmission across classical networks \cite{OSIrefModel} or delivery of end-to-end entangled links by quantum networks \cite{dahlberg_link_2019}.
 
We assume that the underlying network stack is as proposed in \cite{dahlberg_link_2019}. 
In particular, we assume that the link layer is implemented using the protocol in \cite{pompili_experimental_2022}.
This protocol assumes the existence of a time-division multiple-access (TDMA) schedule computed by an external scheduler which determines when entanglement generation can take place. 

Our architecture facilitates the construction of such schedules. We define a periodic timeline for computation and distribution of network schedules to network components. A network schedule is a time ordered plan for the execution of finite duration network tasks. We provide a precise system for translating the demands for service originated by applications running on end nodes into the task executions that make up a schedule. The translation process takes into account the capabilities of the various network components, learned through a capability update process between components and the central controller.

\subsubsection{Network Capabilities}
\label{design considerations: Network capabilities}

The network provides information about its capabilities to the end nodes as a list of pairs $\texttt{(rate, fidelity)}_{i,j}$, describing the rate and fidelity respectively at which entangled links can be generated between a pair of end nodes $i$ and $j$. 
These pairs can also be endowed with extra information about, for example, the expected jitter and the availability of each option. 
We assume that the SDN controller for the network is able to compute these properties from its knowledge about the capabilities of each individual component.

\subsubsection{Timing}\label{subsec:timing-/-time-units} 
The time required for network elements to complete actions can only be estimated with finite precision. At the physical layer, actions have precisely characterised durations, allowing for accurate synchronisation between multiple nodes, with timing precision ranging from tens of picoseconds (ps) to microseconds ($\mu$s), depending on the operations \cite{ThreeNodeQN, donne_design_2024}. Precise timing of a sequence of operations is crucial for processes like entanglement generation \cite{DoubleClickDiamond, humphreys_deterministic_2018, HerEntTrappedIons1, HerEntTrappedIons2, HerEntOriginalAE, HerEntSecondAE, NeutralAtomsHeralded}.

In contrast, at higher layers of the network stack, actions have variable durations and latencies, limiting feasible timing precision to $\mu s$ or milliseconds (ms). For example, transmitting a network schedule over the internet from a central controller to a node 10 km away may take from 50 $\mu$s to several ms. This variability in process duration can be due a variety of sources. Typical examples include that a single process may comprise a large number of computational (low level) operations, that there may be traffic from other processes--creating competition and leading to waiting times for limited computational resources or network bandwidth, or inter-process interactions that necessitate waiting for responses from inter-dependent processes \cite{OSLatency, ModernOS, HardRealTimeComputing}.

To accommodate this variability, our network architecture uses a modular framework, computing and distributing schedules to nodes in advance.

An additional type of timing consideration is that entanglement generation involves sequential non-overlapping attempts, each with a particular sequence of operations, setting a minimum period and maximum rate for the process. This also means operations cannot change mid-attempt without disrupting entanglement generation.

\subsection{Applications}
\label{subsec:quantum-applications}
For simplicity, we assume quantum network applications are executed between two end nodes. However, our architecture is directly compatible with applications involving multiple end nodes.

\subsubsection{Execution Environment}

To execute applications, end nodes require a runtime execution environment.
In this work we assume the Qoala \cite{vecht_qoala_2025} runtime environment is employed, running on QNodeOS \cite{QNodeOS} which has many useful features. 
Qoala breaks applications down into a \textit{program} running on each end node.
Each of these programs is then broken down further into \textit{blocks} of instructions, which can be one of four types: classical local (CL), classical communication (CC), quantum local (QL) or quantum communication (QC).
Quantum communication blocks correspond to generating entangled links. 

When an application is executed, each block of instructions causes a task to be created, which is then scheduled for execution by a local scheduler on that end node. 
The execution of the task corresponds to realising the block of instructions.
Task execution scheduling can either be performed in advance or at runtime. 

The Qoala environment comes equipped with a compiler which runs locally on each node. 
This compiler provides advice about how local hardware parameters (e.g. memory lifetimes) are mapped to requirements on any entangled links produced.
The compiler also produces metadata for every block and for an entire program.
This metadata covers any constraints on the execution of tasks corresponding to the constituent blocks of a program. 
This metadata is used in a process called \textit{capability negotiation} during which the nodes ensure that the execution of the various programs comprising the application is compatible. 

Whilst here we assume use of the Qoala environment, our architecture is compatible with any runtime environment with an equivalent notion of task scheduling and a compiler that provides equivalent metadata.

\subsubsection{Application Classes}
\label{subsec:DesignConsiderations:ApplicationClasses}
There are many different quantum network applications, and each will have different requirements for entangled links generated by the network.
However, they may be broadly grouped into two different classes, \textit{measure-directly} and \textit{create-and-keep} \cite{dahlberg_link_2019}.
In measure-directly (MD) applications, qubits are measured as soon as entanglement is produced, and no states are kept in memory.
In this case, the demand for the generation of entangled links is elastic \cite{shenker_fundamental_1995}.
Examples of MD applications include \textit{quantum key distribution} (QKD), which facilitates provably secure secret sharing \cite{BB84, ekert_quantum_1991}, and \textit{deterministic teleportation} \cite{unconditionalTeleportation, determinsticTeleporationAtoms}. 

In create-and-keep (CK) applications, qubits are stored in memory after entangled links have been generated.
Typically CK applications will require many qubits to be stored in memory simultaneously.
This places limitations on the spacing between the generation of new links. 
An example of a CK application is \textit{blind quantum computing} (BQC), which facilitates secure remote computation  (e.g. \cite{bqc1, bqc2, childs_secure_2005}).

\subsubsection{Architecture}\label{subsec:node-architecture}
Our architecture is agnostic to the architecture of the end nodes of the network. 
However, for convenience we assume that processing nodes have a single monolithic \textit{quantum processing unit} (QPU), which is used both for entanglement generation and performing local gates and measurements.
Such a QPU is assumed to be only capable of performing a single operation at any given time.
This is in line with state-of-the-art implementations of end nodes \cite{pompili_experimental_2022, HerEntTrappedIons2}.

We do not make any assumptions about the classical processing capabilities of the end nodes.

\subsubsection{Knowledge of the network}
We assume that an end node has minimal knowledge about the rest of the network. 
In particular, we assume that \textit{a priori} an end node only knows the capabilities and status of their own hardware, and the identities of any neighbouring network components and the identity of any control devices in the network.

Furthermore, we assume that an end node only knows the programs which are running on itself.
In particular, it is not required that an end node \textit{a priori} knows the precise design of the program on the other node in the application.
Additionally, each program should be independent of any other, and the ability to execute a program should not require knowledge of the other programs running on the end node.

\section{Network Architecture}\label{sec: Network Architecture}
\begin{figure*}[t]
    \centering

    \begin{tikzpicture}[
            node distance = 1cm,
            action/.style={rectangle,
                            draw=black!60,
                            fill=white,
                            very thick,
                            align=center,
                            minimum width=2cm,
                            minimum height = 1.5cm,
                            },
            decision/.style={diamond,
                            draw=black!60,
                            fill=white,
                            very thick,
                            align=center,
                            minimum width=2cm,
                            minimum height = 2cm,
                            aspect=2,
                            },
            A/.style={->,thick}
]

    \node[action] (NCU) {Network\\Capability\\Update (\ref{ssec: Arch - NCU})};
    \node[action] (CN) [right=of NCU] {Capability\\Negotiation\\\eqref{ssec: Arch - CN}};
    \node[action] (DS) [right=of CN] {Demand\\Submission\\(\ref{subsec:arch---demand-format}, \ref{ssec:Arch - Demand Reg})};
    \node[action] (NS) [right=of DS] {Network\\Scheduling\\\eqref{ssec: Arch - NS}};
    \node[action] (NdS) [right=of NS] {Schedule\\Dsitribution\\\eqref{subsec: Arch - distribution}};

    \draw[A] (NCU) -- (CN);
    \draw[A] (CN) -- (DS);
    \draw[A] (DS) -- (NS);
    \draw[A] (NS) -- (NdS);

\end{tikzpicture}

     \caption[Flow of information through the stages of the architecture]{Flow of information through the stages of the architecture. The processes of `Network Capability Update' and `Capability Negotiation' allow the nodes to gather enough information to be able to submit a unified demand. These demands are then used to construct a central network schedule, which in turn is used when constructing the local node schedules. }
    \label{fig:Information Flow}
\end{figure*}
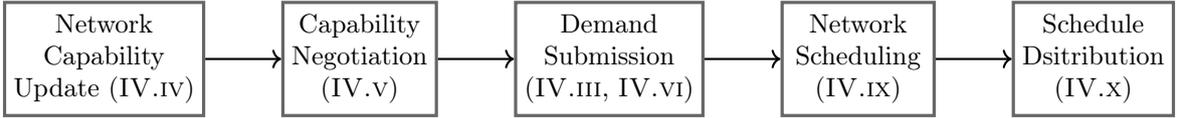

Our proposed network architecture is broken up into five stages, each of which corresponds to a distinct required process. 
These stages are: \textit{network capability update} and \textit{capability negotiation}, where the end nodes learn information about each other and the network in order to construct a unified demand; \textit{demand submission}, where end nodes submit their demands to the central controller; and finally \textit{network scheduling} and \textit{schedule distribution}, where the central controller computes the network schedule before distributing it to the end nodes. 
The stages of the architecture are summarised in Figure~\ref{fig:Information Flow}.

Our architecture introduces several unique features which ensure compatibility of the architecture with end nodes and program execution. 
Firstly, we periodically distribute network schedules to end nodes well before the start time of the network schedule, which allows nodes enough time to compute local schedules that incorporate the network schedule. 
Secondly, we introduce a demand format which takes into account local processing time requirements.
This demand format includes a precise description of the entangled links which are required by an application, through a notion of a \textit{packet of entanglement}. 
Finally, the format and construction of our network schedules ensure that end nodes are certain about when they can attempt to generate entanglement, allowing for efficient and effective program execution. 

In the rest of this section, we first introduce several preliminaries which precisely define how an application's requirements translate to a demand for service from end nodes to the network. Then, we provide a detailed description of each stage of the network architecture. 

\subsection{Periodic Computation and Distribution of Schedules}
\label{ssec: Net Arch: Schedule Comp and Dist}

\begin{figure*}
    \centering
    \begin{tikzpicture}
[
			node distance = 1cm,
            background rectangle/.style={fill=white},
            collect demands/.style={rectangle,
                            draw=blue!60, 
                            fill=blue!20, 
                            very thick, 
                            align=center, 
                            minimum width=8.5cm, 
                            minimum height = 6.5cm,
                            rounded corners,
                            },
            Component/.style={rectangle,
                            draw=red!60, 
                            fill=red!20, 
                            very thick, 
                            align=center, 
                            minimum width=3cm, 
                            minimum height = 1cm,
                            rounded corners,
                            },
            PU/.style={rectangle,
                            draw=red!60, 
                            fill=red!20, 
                            very thick, 
                            align=center, 
                            minimum width=1cm, 
                            minimum height = 1cm,
                            rounded corners,
                            },
            QNtwkCmp/.style={rectangle,
                            draw=Fuchsia!60, 
                            fill=Fuchsia!20, 
                            very thick, 
                            align=center, 
                            minimum width=3.3cm, 
                            minimum height = 0.5cm,
                            },
            CNtwkCmp/.style={rectangle,
                            draw=BlueGreen!60, 
                            fill=BlueGreen!20, 
                            very thick, 
                            align=center, 
                            minimum width=3cm,  
                            minimum height = 0.5cm,
                            },
            CC/.style={rectangle,
                            draw=orange!60, 
                            fill=orange!20, 
                            very thick, 
                            align=center, 
                            minimum width=20cm, 
                            minimum height = 5cm,
                            rounded corners,
                            },
            SA/.style={thick, ->,},
            DA/.style={thick, <->,},
]
   
\draw[thick] (-9.275,3.5) --  (-9.275,-0.5) node [below] {$k-1$} ;
\draw[thick] (-6.175,3.5) --  (-6.175,-0.5) node [below] {$k$} ;
\draw (-3.075, 3.5) --  (-3.075,-0.5) node [below] {$k+1$} [thick];
\draw (0.025,3.5) --  (0.025,-0.5) node [below] {$k+2$} [thick];
\draw[->, thick] (-9.35,-0.4) -- (0.4,-0.4); 

\newcommand{\actionStack}[6]{
    \coordinate (root) at #3;
    \ifnum#1=1
        
\node[CNtwkCmp] (execute schedule) [left=0cm of root, minimum width=3cm, preaction={fill, Blue!20}, pattern=#2, pattern color=Blue!60, draw=Blue!60] {};
        \node[CNtwkCmp] (compute node schedule) [left=0cm of execute schedule, yshift=0.6cm, xshift = -0.075cm, minimum width = 0.5cm, draw=YellowOrange!60, preaction={fill, YellowOrange!20}, pattern=#2, pattern color=YellowOrange!60] {};
        \node[CNtwkCmp] (distribute schedule) [left=-0cm of compute node schedule, yshift=0.6cm, xshift = 0cm, minimum width = 1.0cm, draw=Red!60, preaction={fill, Red!20}, pattern=#2, pattern color=Red!60] {};
        \node[CNtwkCmp] (compute schedule) [left=0cm of distribute schedule, xshift=0cm, yshift=0.6cm, minimum width = 1.1cm, preaction={fill, Green!20}, draw=Green!60, pattern=#2, pattern color=ForestGreen!60] {};
        \node[CNtwkCmp] (admit demands) [left=-0cm of compute schedule, yshift=0.6cm, minimum width = 0.27cm, preaction={fill, Green!20}, draw=Green!60, pattern=#2, pattern color=ForestGreen!60] {};
        \node[CNtwkCmp] (collect demands) [left=0cm of admit demands, yshift=0.6cm, minimum width = 3cm, xshift = -0.05cm,  draw = Fuchsia!60, very thick, fill=Fuchsia!20, pattern=#2, pattern color=Fuchsia!60, preaction={fill, Fuchsia!20}]{};        
    \else
\ifnum#4=1
            \node[CNtwkCmp] (execute schedule prev) [draw=Black!20, minimum width=3cm, left=0cm of root, preaction={fill, Black!5}, pattern=#2, pattern color=Black!20] {};
        \fi
        \ifnum#5=1
            \node[CNtwkCmp] (compute node schedule) [left=3.025cm of root, yshift=0.6cm, xshift = -0.075cm, minimum width = 0.5cm, draw=Black!20, preaction={fill, Black!5}, pattern=#2, pattern color=Black!20] {};
            \node[CNtwkCmp] (distribute schedule) [left=0cm of compute node schedule, yshift=0.6cm, xshift = -0cm, minimum width = 1.0cm, draw=Black!20, preaction={fill, Black!5}, pattern=#2, pattern color=Black!20] {};
            \node[CNtwkCmp] (compute schedule) [left=0cm of distribute schedule, xshift=0cm, yshift=0.6cm, minimum width = 1.1cm, preaction={fill, Black!5}, draw=Black!20, pattern=#2, pattern color=Black!20] {};
            \node[CNtwkCmp] (admit demands) [left=-0cm of compute schedule, yshift=0.6cm, minimum width = 0.27cm, preaction={fill, Black!5},    draw=Black!20, pattern=#2, pattern color=Black!20] {};        
        \fi
        \ifnum#6=1
            \node[CNtwkCmp] (collect demands) [left=6.15cm of root, yshift=3cm, minimum width = 3cm, xshift = -0.05cm,  draw = Black!20, very thick, fill=Black!20, pattern=#2, pattern color=Black!20, preaction={fill, Black!5}]{}; 
        \fi
    \fi
}

\actionStack{0}{none}{(6.2, 0)}001
\actionStack{0}{none}{(3.1, 0)}011
\actionStack{1}{crosshatch}{(0,0)}111
\actionStack{0}{none}{(-3.1, 0)}110
\actionStack{0}{none}{(-6.2, 0)}100

\coordinate (label root) at (-9.35,-0.4);

\node[align=right] at ($(label root|-collect demands)$) [left]  {Demand Registration};
\node[align=right] at ($(label root|-admit demands)$) [left] {Scheduler Admission Control};
\node[align=right] at ($(label root|-compute schedule)$) [left]  {Compute Schedule};
\node[align=right] at ($(label root|-distribute schedule)$) [left]  {Distribute Schedule};
\node[align=right] at ($(label root|-execute schedule)$) [left]  {Execute Schedule};
\node[align=left] at ($(label root|-compute node schedule)$) [left] {Compute Node Schedule};

\node[align=center] at ($(-9.35,-1)!0.5!(0.4, -1)$) [below] {Scheduling Interval};

\end{tikzpicture}     \caption[Example of the timings for computing, distributing and executing the $k$th network schedule.]{Example of the timings for computing, distributing and executing the $k$th network schedule (in colour and hatched). 
    In grey and unhatched are the corresponding operations for preceding and subsequent network schedules.
    The process of registering received demands is continuous, however we indicate here the time during which if a demand is submitted, then it will be first considered for scheduler admission as part of computing the $k$th network schedule.}
    \label{fig:network scheduling timings}
\end{figure*}
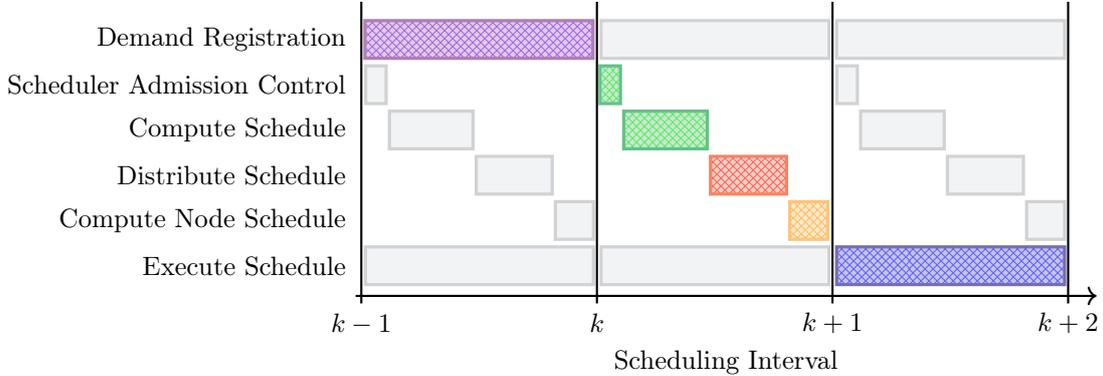

The essential task of computing and distributing a network schedule is executed periodically by the central controller.
Each schedule is associated with a version identifier and covers an identical execution time known as the \textit{scheduling interval} (denoted $T_{SI}$).
Periodic triggers for computation and distribution of the schedule are defined with respect to the $T_{SI}$. 
Figure~\ref{fig:network scheduling timings} illustrates possible trigger timings for the components of this periodic task
and illustrates the back-to-back execution of subsequent schedules.

The primary advantage of a periodic approach to scheduling, as compared to various possible on-line approaches to scheduling, is that it limits the number of updates to the schedule that may be triggered. In particular, the central controller does not need to update the schedule every time a new demand is submitted, which reduces the number of time-consuming interruptions during scheduling.
This approach is also advantageous for end nodes, which receive only finalised network schedules. Consequently,
unnecessary interruptions to local schedules and program execution are avoided. Moreover, any buffer between the time a schedule is received and the time it takes effect allows end nodes to optimise their local program schedules without compromising execution of the network schedule.

Additionally, as compared to on-line network scheduling with an unbounded number of updates, the periodic distribution of pre-computed network schedules reduces the number of messages which need to be sent by the central controller, serving to improve the reliability of the network. 
With this approach to scheduling, all the components of the network know a) when the schedule is changing and b) which schedule version should be in effect when, reducing the amount of communication required between the end nodes and the central controller.
These features reduce the likelihood that network components attempt to execute different versions of the network schedule.

\subsection{Application Sessions}\label{ssec: Net Arch: application-structure}

As the outcome of many quantum applications is probabilistic, it is often required to execute the same application many times to extract a useful and reliable output.
We refer to each of these individual executions of an application as an \textit{application instance}.

It is also often the case that the nodes will require that all these instances are executed before some time elapses. 
For example, suppose Alice and Bob wish to communicate securely within the next two hours.
To do so, they may use QKD to generate a raw key and extract a secret key which can then be used to communicate. To generate enough raw key would then require them to execute many instances of the QKD application during the next two hours.

To capture these requirements, we define an \textit{application session}. 
\begin{define}[Application Session]
Suppose end nodes $\mathcal{N}=(\texttt{node1, node2,...})$ wish to execute application \texttt{App} at least $\ninst$ times, before time $\texpiry$.

Then we write the corresponding \textit{application session} as 
\begin{equation}
    \mathcal S = (\mathcal{N}, \texttt{App},\ninst, \texpiry) \label{eq: maths defn of session}
\end{equation}
\end{define}

It is not required that the instances are all identically executed.
For example in an application such as verifiable blind quantum computing, (e.g. \cite{leichtle_verifying_2021}), some instances may correspond to computation rounds (where the calculation is performed) and others to test rounds (where the veracity of the server is checked).
However, all instances should use the same programs on each end node. 

Furthermore, we expect that the  specified expiry time, $\texpiry$, will be much longer than any time-scale of the network.
For instance, if the network updates (see \S\ref{ssec: Arch - NS}) on the order of minutes, we expect $\texpiry$ to be on the order of hours.

When pairs of nodes send their demands for entanglement to be generated to the network, they will do so on a per-session basis, rather than for each individual instance. 
This reduces the number of demands that the central controller has to process.
Furthermore, it gives the central controller an overview of the scope of a demand, and therefore it can make better decisions regarding whether and how to give it service.

\subsubsection{Executing Application Instances and Obtaining Minimal Service}
\label{subsubsec: Arx - Sessions - Execution Defns}
It will be useful to be able to talk about whether program instances have executed successfully.
However, because of the imperfect nature of qubits, even if all gates are performed perfectly, an `incorrect' outcome may be obtained, which could be interpreted as an unsuccessful execution. 
Therefore, for clarity about what we mean when we talk about application instances successfully executing, we make the following series of definitions:
\begin{define}[Application Instance Execution]
    We say an instance of an application is \textit{executed} when at least one of the blocks of instructions are performed by the relevant processor.
\end{define}
\begin{define}[Successful Quantum Block Execution]
    We say a block consisting of local gates \textit{executed successfully} if all included gates were applied to the qubits without destroying them and a signal confirming this is passed back to the local scheduler. We say a quantum communication (entanglement generation) block \textit{executed successfully} if a \textit{packet of entanglement} (see \S\ref{subsubsec: Arx - Demand Formant - Packets}) is generated, and a signal confirming this is passed to the local scheduler.
\end{define}
\begin{define}[Successful Application Instance Execution]
    We say an application instance \textit{executed successfully} if all blocks are successfully executed, including receiving confirmation signals.
\end{define}
\begin{define}[Quantum Successful Execution]
    We say an application instance achieved \textit{quantum success} if it was successfully executed and the correct outcome was obtained. 
\end{define}

\begin{define}[Minimal Service]
    We say a session obtains \textit{minimal service} in the event of $\ninst$ successfully executed instances of the application before the session expiry time $\texpiry$ elapses. \label{def: minimal service for a session}
\end{define}

The aim of the network is to ensure that as many sessions as possible obtain minimal service.
Note that in the literature, what we refer to as `quantum successful execution' is commonly referred to as ``application success'', for example, as in \cite{vecht_qoala_2025}. However, it is useful to make the distinction between an instance successfully executing and an instance achieving quantum success.
In particular, the network aims to ensure sessions obtain minimal service.
To do so, it facilitates the successful execution of application instances by scheduling time to create packets of entanglement.
In contrast, the aim of end nodes is to achieve quantum successes from successfully executed instances by suitably scheduling the execution of local gates and operations.

We also define a notion of `load' on the network, corresponding to the number of pairs of users which have submitted demands for entanglement generation.
In particular, if the load on the network is high, this should correspond to a decrease in the likelihood of a session obtaining minimal service from the network, due to the network resources not being able to serve all submitted demands simultaneously. 

\subsection{Demand Format}\label{subsec:arch---demand-format}

In order to schedule time for the generation of entangled links the central controller requires two sets of information. Firstly, it needs to know the quantity and quality of entangled links that must be produced. Secondly, it requires knowledge of the frequency with which and over what time period these entangled links must be produced. 
Therefore, the demand format needs to capture all this information.
Note that the application session does not capture all of this information, so we cannot simply use the application session as the demand submitted to the network.
Furthermore, the network should be agnostic to the specific application being run and only be aware of the desired entangled links to be generated.

\subsubsection{Packets of Entanglement}\label{subsubsec: Arx - Demand Formant - Packets}
We first address specification of the quality and quantity of entanglement required. 
Recall from \S\ref{subsec:DesignConsiderations:ApplicationClasses} that when executing an application from the create-and-keep class, each instance will require many co-existing entangled links.
Each of these links cannot be too old because quantum memories, which store them, exhibit time-dependent decoherence, causing the quality of the links to decrease over time.
The maximum age of a link depends both on the hardware of the specific end nodes and the initial fidelity of the entangled link.
One approach to satisfy these requirements is to strictly control the jitter between generated entangled links, for example as in \cite{skrzypczyk_architecture_2021}.
However, we instead consider the generation of \textit{packets of entanglement}, rather than individual links. 
The generation of a packet of entanglement corresponds to the co-existence of all required entangled links for an instance of an application.
There is no time constraint between the generation of each packet of entanglement, as each instance of an application is independent.
Therefore, following the definitions in \cite{shenker_fundamental_1995}, any demands, both from sessions of measure-directly and of create-and-keep applications can be treated as elastic.  

One may also attempt to control the jitter between the generation of packets, though the ability to do so will depend on the network scheduling algorithm employed.
Furthermore, this should not be required in order for a network application to be executed.  

The following definitions, motivated by the formalism set out in \cite{davies_tools_2024}, precisely specify what is meant by packets of entanglement.

\begin{define}[Window]
    A \textit{window} in the context of entanglement generation is the longest time that an entangled link can be kept in memory without decohering too much to be useful.
    If a link has been in memory for longer than the length of the window it is discarded.
    In particular, this means that all the entangled links required to execute an instance of a quantum application need to be generated within a time window's duration of each other. 
\end{define}

\begin{define}[Entanglement Packet]
      A \textit{packet of entanglement} or \textit{entanglement packet} for a given application session is the tuple $(w,s,F_{\min})$, where $w$ is the time window within which all entangled links must be created, $s$ is the required number of entangled links and $F_{\min}$ is the minimum fidelity 
new links may be created with. 
\end{define}

To see how this is useful, suppose Alice and Bob wish to execute two instances of an application, each requiring three entangled links.
However, Alice and Bob's hardware is such that each entangled link can only be stored for 0.5s.
Instead of saying that Alice and Bob require 6 entangled links with expected fidelity $F$, it is much more precise to say they need two $(0.5s, 3, F)$ packets of entanglement.
In particular, if the network allocated resources such that Alice and Bob sequentially generated 6 entangled links, each 1 second apart from the other, this would satisfy the former requirement, but Alice and Bob would still not be able to execute their applications successfully. 
By specifying the packet of entanglement required, the network can allocate resources for sufficiently long periods in the network schedule that the required entanglement, i.e. a packet, can be produced. The packet formalism thereby ensures that application instances can actually be executed.

\begin{define}[Packet Suitability]
    We say that a packet of entanglement is \textit{suitable} for an application if the existence of entangled links adhering to the form of the packet would allow an instance of the application to be executed with an acceptable probability of achieving quantum success. 

\end{define}

It is possible for an application instance to have multiple suitable packets of entanglement.
This may arise from, for example, accepting links which are generated with a lower expected fidelity and shortening the window to compensate.
The end nodes could also include extra links which may not be directly required by the protocol. 

End nodes include a finite subset of all possible suitable packets in their demand.
We write $\mathcal{P}$ for the set of suitable packets included in the demand.

In order to compute the set of suitable packets, the nodes need to know what quality of entanglement the network can produce, as well as the hardware on the other node(s). 
These pieces of information are obtained in the \textit{network capability update} and \textit{capability negotiation} phases respectively. Given this, they can establish how long the links can be stored in memory, and thus construct the set of suitable packets.

\subsubsection{Timing Constraints}
We now address communicating the frequency and time-period over which packets of entanglement are produced. 
For each suitable packet of entanglement, the nodes specify a average rate $R$ at which they wish such packets to be produced.
This rate can be set to 0, which indicates to the central controller that the nodes will be satisfied with the minimum possible rate of packet generation which almost certainly ensures the session obtains minimal service. 
In this case, the nodes also submit $N_{inst}$ as part of the demand, so the network can calculate this minimum rate when the demand is accepted for scheduling.

The central controller will define a \textit{service model} which specifies how the requested rate will be treated. 
For example, the rate may be met exactly, or alternatively the network may increase or decrease the rate at which packets are generated within specified limits, depending on the current network load.

Alongside the packets and associated rates, the nodes also submit two further pieces of timing information. 
Firstly, they include the expiry time of the session $\texpiry$.
Secondly, they include a minimum separation between attempts to generate packets of entanglement. 
This minimum separation is included to ensure that there is sufficient time for local operations to be performed before the next allocated period of time for generating entangled links begins. Local operations may either be additional blocks of quantum operations in the application program or operations to reset the hardware between subsequent attempts to generate a packet.

Submitted rates may depend on factors such as how often the nodes intend to execute instances of an application, as well as pre-existing device agreements with the network.
Thus, the values are determined as part of the \textit{capability negotiation} phase.
Determination of the minimum separation required between attempts to generation a packet and an appropriate expiry time also requires input from both nodes, thus these values are set during \textit{capability negotiation}.

\subsubsection{Full Demand}

Combining all of the above is the demand which end nodes submit to the network:

\begin{align}
    \mathcal{D} = \left(\left\lbrace\left(w,s,F,R\right)_p\right\rbrace_{p\in \mathcal{P}}; t_\texttt{minsep}, t_\texttt{expiry}; \ninst\right),\label{eq: Demand}
\end{align} where $\mathcal{P}$ is the set of submitted  suitable packets.

As the demand format needs to be compatible with all end nodes, the time dependent parameters $w,R_\texttt{packet},t_\texttt{minsep}$ and $t_\texttt{expiry}$ should be specified in terms of real-time units, such as seconds or per-second as applicable, rather than in terms of local or network time slots.
The central controller can then convert them to a notion of time-slots, if required, when computing and distributing the schedule.

\subsection{Network Capabilities Update}\label{ssec: Arch - NCU}

To enable construction of the suitable packets of entanglement, firstly the end nodes need to know whether the network can generate entangled links between them, and secondly at what quality these links can be produced. 
Providing the nodes with this information is the primary objective of the \textit{network capabilities update} phase of our architecture (interaction \textbf{A} in Figure \ref{fig:Network Interactions}). 

The application stack obtains this information in the following manner. 
Firstly, a query is sent to the quantum network agent (QNA) (see Figure~\ref{fig:Network Interactions}).
This query can be either for an overview of the network, or for the specific entanglement generation capabilities with another party. 
If the QNA has recently obtained the requested information, then it responds directly, otherwise the query is forwarded to the \textit{network capabilities manager} of the central controller. 

In the case of an overview request, the response includes with whom entanglement can be generated, as well as general information about the status of the network such as the current load. 
If the application stack requests the capabilities for a specific node, 
this information is returned in the form of $(R,F)$ rate-fidelity pairs describing the rate at which end-to-end links can be generated with fidelity $F$. 

In order to determine the quality of entanglement which can be generated, the central controller performs the following tasks:
First, it establishes along which paths through the network entangled links can be generated. 
Then, along these paths the central controller devises a scheme which will enable end-to-end entangled links to be produced. 
Using the information which the central controller has about the fidelity of the entangled links which can be produced along each of the segments of this path, the overall end-to-end fidelity can be determined and communicated back to the requesting end node's quantum network agent.

The central controller retains the right to be selective about which possibilities it communicates. 
For instance, it may discount certain paths or configurations which would put excessive pressure on the network. 
Furthermore, the network only communicates the quality of entanglement and \textit{not} the paths back to the nodes. 
This is to give the controller flexibility, where possible, to create the same quality links using multiple different paths as well as to maintain the agnosticism of the end nodes as to the internals of the network.

\subsection{Capability Negotiation} \label{ssec: Arch - CN}

Before any application can be run, or a demand submitted to the network, the nodes must align amongst themselves exactly how the application will be run.
Ultimately the aim of this is to create an application session and corresponding demand, as well as finalising any metadata about the programs which still needs to be set. 

In order to do this, the nodes carry out \textit{capability negotiation} (interaction \textbf{B} in Figure~\ref{fig:Network Interactions}).
During this phase the nodes exchange relevant information, for example the quantity and quality of qubits which can be made available and how other end nodes need to interact with their programs. 
In particular, combined with the data from network capability update, the nodes should be able to exchange sufficient information to determine the acceptable packets and calculate/decide upon the packet generation rates they will request from the network. 
Furthermore, by the end of this exchange, the nodes will have decided upon the values of $\ninst, \texpiry$ and $t_{\texttt{minsep}}$.

Therefore, once capability negotiation has concluded, up to classical communication in the application itself, the end nodes should be able to execute their programs independently and without further interaction between them. 

\subsection{Demand Registration} \label{ssec:Arch - Demand Reg}

When demands are received by the central controller (interaction \textbf{C} in Figure~\ref{fig:Network Interactions}), they undergo an initial registration process.
If a demand passes this process, then it is placed into the \textit{demand queue} for consideration by the scheduler admission control. 
Otherwise, it is immediately rejected by the network. 
The end node which submitted the demand is informed of the outcome of the demand registration process, and additionally can be informed as to the reasons for rejection. 
In the case no such acknowledgement arrives, end nodes should assume that the demand has been rejected. 

The main aim of this process is to filter out any obviously infeasible or unreasonable demands.
However, on top of this, this process may also reject demands based on the load which the network is experiencing. 
For example, if load on the network is high, then the demand registration process may also immediately reject demands with a high expected queuing time. 
The precise rules which the demand registration process implements will depend on the network implementation and the types of behaviour which the network operator desires.

\subsubsection{Leaving the Demand Queue}
Once a demand has been registered and placed into the demand queue, there are a few ways in which it can be removed from the queue. 
The positive outcome for a demand is that it passes the scheduler admission control, and is accepted to be scheduled. 
Depending on the load on the network and the nature of the demand, this may happen at the start of the next scheduling interval. Alternatively, a demand may possibly be held in the queue for several scheduling intervals, until there is sufficient capacity for the network to serve it. 

The other main reason that a demand can be removed from the queue is if that demand expires. 
However, at the discretion of the central controller, demands may also be removed from the queue before they expire if such removal benefits overall the overall performance of the network. 
For example, one such rule may be that any demand which would have failed demand registration had it been submitted at the current time is removed from the demand queue. 
As with demand registration, the exact rules governing premature removal from the demand queue should be specified in an implementation.

\subsection{Session Initialisation}
\begin{figure*}[t]
    \centering
    \resizebox{17cm}{!}{

\usetikzlibrary {backgrounds}
\tikzset{background grid/.style={thick,draw=red,step=.5cm}}
\begin{tikzpicture}[
			node distance = 1cm,
            background rectangle/.style={fill=white},
            endNode/.style={rectangle,
                            draw=blue!60, 
                            fill=blue!20, 
                            very thick, 
                            align=center, 
                            minimum width=8.5cm, 
                            minimum height = 6.5cm,
                            rounded corners,
                            },
            Component/.style={rectangle,
                            draw=red!60, 
                            fill=red!20, 
                            very thick, 
                            align=center, 
                            minimum width=3cm, 
                            minimum height = 1cm,
                            rounded corners,
                            },
            PU/.style={rectangle,
                            draw=ProcessBlue!60, 
                            fill=ProcessBlue!20, 
                            very thick, 
                            align=center, 
                            minimum width=1cm, 
                            minimum height = 1cm,
                            rounded corners,
                            },
            QNtwkCmp/.style={rectangle,
                            draw=Fuchsia!60, 
                            fill=Fuchsia!20, 
                            very thick, 
                            align=center, 
                            minimum width=3cm, 
                            minimum height = 1cm,
                            rounded corners,
                            },
            obj/.style={rectangle,
                            draw=OliveGreen!60, 
                            fill=OliveGreen!20, 
                            very thick, 
                            align=center, 
                            minimum width=3cm, 
                            minimum height = 1cm,
                            rounded corners,
                            },
            CC/.style={rectangle,
                            draw=orange!60, 
                            fill=orange!20, 
                            very thick, 
                            align=center, 
                            minimum width=20cm, 
                            minimum height = 3cm,
                            rounded corners,
                            },
            SA/.style={thick, ->},
            DA/.style={thick, <->},
            SA1/.style={thick, <-},
]

\node (root) at (0,0){};

\node[endNode] (A) [left=1.5 of root] {}; 
\node[] (A_label) [below=0 of A] {End Node A};

\node[QNtwkCmp, rounded corners, minimum width=7.5cm,minimum height=1cm] (A_QNAgent) [above=1.85 of A.center] {Quantum Network Agent};

\node[obj] (A_NS) [left=0.75 of A.center, yshift=0.75cm] {Network\\Schedule};
\node[Component] (A_LS) [below=0.5 of A_NS] {Node Schedule\\Computation};
\node[PU] (A_QPU) [below =0.5 of A_LS, xshift=1cm]{QPS};
\node[PU] (A_CPU) [below =0.5 of A_LS, xshift=-1cm]{CPS};

\node[Component] (A_AS) [right=1.25 of A.center, yshift=0.75cm, minimum width=2cm] {Application\\Stack};
\node[QNtwkCmp] (A_QNS) [below = 1.625 of A_AS, xshift=-0.025cm] {Quantum\\Network\\Stack};

\draw[DA, double] (A_QPU.east) -- ($(A_QNS.west |- A_QPU)$);
\draw[DA] (A_QPU.west) -- (A_CPU.east);

\draw[DA] (A_QPU.north) -- ($(A_LS.south)+(1,0)$);
\draw[DA] (A_CPU.north) -- ($(A_LS.south)+(-1,0)$);

\draw[SA] (A_NS.south) -- (A_LS.north);
\draw[SA] ($(A_NS.north |- A_QNAgent.south)$)  -- (A_NS.north);
\draw[DA] ($(A_AS.north |- A_QNAgent.south)$)  -- (A_AS.north) node [midway, right] {\textbf{A}};

\coordinate (Ax) at ($(A_QNS.west)!0.5!(A_QNS.north west)$);
\coordinate (Ay) at ($(A_AS.south)!0.25!(A_AS.south east)$);
\coordinate (Az) at ($(A_QNS.north)!0.66!(A_QNS.north west)$);

\draw[SA] ($(A_LS.north east)!0.67!(A_LS.south east)$) -| ($(A.center |- Ax)$) -- (Ax);
\draw[SA1] ($(A_LS.north east)!0.33!(A_LS.south east)$) -| ($(A_AS.south)!0.5!(A_AS.south west)$)  node [midway, right] {\textbf{D}};
\draw[SA] (Ay) -- ($(A_QNS.north -| Ay)$) node [midway, right] {\textbf{C,D}};
\draw[DA] (Az) -- ($(Az |- A.center)$) -| (A_QNAgent.south) node [midway, above, xshift=0.25cm] {\textbf{C}};

\node[endNode] (B) [right=1.5 of root] {}; 
\node[] (B_label) [below=0 of B] {End Node B};

\node[QNtwkCmp, rounded corners, minimum width=7.5cm,minimum height=1cm] (B_QNAgent) [above=1.85 of B.center] {Quantum Network Agent};

\node[obj] (B_NS) [right=0.75 of B.center, yshift=0.75cm] {Network\\Schedule};
\node[Component] (B_LS) [below=0.5 of B_NS] {Node Schedule\\Computation};
\node[PU] (B_QPU) [below =0.5 of B_LS, xshift=-1cm]{QPS};
\node[PU] (B_CPU) [below =0.5 of B_LS, xshift=1cm]{CPS};

\node[Component, minimum width = 2cm] (B_AS) [left=1.25 of B.center, yshift=0.75cm] {Application\\Stack};

\node[QNtwkCmp] (B_QNS) [below = 1.625 of B_AS, xshift=0.025cm] {Quantum\\Network\\Stack};

\draw[DA, double] (B_QPU.west) -- ($(B_QNS.east |- B_QPU)$);

\draw[DA] (B_QPU.east) -- (B_CPU.west);

\draw[DA] (B_QPU.north) -- ($(B_QPU.north |- B_LS.south)$) node [midway, left] {\textbf{G}}; 
\draw[DA] (B_CPU.north) -- ($(B_CPU.north |- B_LS.south)$) node [midway, left] {\textbf{G}};

\draw[SA] (B_NS.south) -- (B_LS.north) node [midway, left] {\textbf{F}};
\draw[SA] ($(B_NS.north |- B_QNAgent.south)$)  -- (B_NS.north) node [midway, left] {\textbf{F}};
\draw[DA] ($(B_AS.north |- B_QNAgent.south)$)  -- (B_AS.north);

\coordinate (Bx) at ($(B_QNS.east)!0.5!(B_QNS.north east)$);
\coordinate (By) at ($(B_AS.south)!0.25!(B_AS.south west)$);
\coordinate (Bz) at ($(B_QNS.north)!0.66!(B_QNS.north east)$);

\draw[SA] ($(B_LS.north west)!0.67!(B_LS.south west)$) -| ($(B.center |- Bx)$) -- (Bx) node [midway, above] {\textbf{G}};
\draw[SA1] ($(B_LS.north west)!0.33!(B_LS.south west)$) -| ($(B_AS.south)!0.5!(B_AS.south east)$);
\draw[SA] (By) -- ($(B_QNS.north -| By)$);
\draw[DA] (Bz) -- ($(Bz |- B.center)$) -| (B_QNAgent.south);

\node[CC, minimum height = 4cm] (CC) [above = 4 of root]{};

\node[Component, minimum width = 5cm] (CC_NCM) [above=0.5 of CC.south] {Network Capabilities Manager};
\node (CC_label) [above=0 of CC] {SDN (Centralised) Controller};

\node[Component] (CC_PGTC) [below=0.5 of CC.north] {PGT\\Creation};
\node[Component] (CC_DQ) [left=0.75 of CC_PGTC] {Scheduler\\Admission};
\node[Component] (CC_DR) [left=0.75 of CC_DQ] {Demand\\Registration};

\node[Component] (CC_NS) [right=0.75 of CC_PGTC] {Network Schedule\\Computation};
\node[obj] (CC_NS_obj) [right=0.75 of CC_NS] {Network\\Schedule};

\draw[SA] (CC_DR) -- (CC_DQ);
\draw[SA] (CC_DQ) -- (CC_PGTC);
\draw[SA] (CC_PGTC) -- (CC_NS);
\draw[SA] (CC_NS) -- (CC_NS_obj);

\draw[dotted, thick] ($(CC_DR.south west) + (-0.2,-0.2)$) rectangle ($(CC_NS_obj.north east) + (0.2,0.2)$);

\coordinate (CCx) at ($(CC_NCM.west -| A_QNAgent.north)$);

\draw[DA] (CC_DR) |- (CCx) node [midway, above, xshift=0.25cm] {\textbf{C}}  -- (A_QNAgent.north);
\draw[SA] (CC_NS_obj.south) |- ($(B_NS.north |- CC_NCM.east)$)  node [midway, above, xshift=0.25cm] {\textbf{E}} -- ($(B_NS.north |- B_QNAgent.north)$);

\draw[DA] ($(A_QNAgent.north -| A_AS.north)$) |- (CC_NCM) node [midway, left] {\textbf{A}};
\draw[DA] ($(B_QNAgent.north -| B_AS.north)$) |- (CC_NCM);

\draw[DA] (A_AS) -- (B_AS) node [midway, above] {\textbf{B}};

\draw[SA, double] ($(root |- A_QNS.east) - (0.5, 0)$) -- (A_QNS.east);
\draw[SA, double] ($(root |- B_QNS.west) + (0.5, 0)$) -- (B_QNS.west);

\node[] at ($(root |- A_QNS.east)$) {...};

\coordinate (Aa) at ($(A_NS.north |- A_QNAgent.north)$);
\coordinate (Ab) at ($(Aa |- CC.south)$);
\coordinate (Ac) at ($(Aa)!0.4!(Ab)$);
\coordinate (Ad) at ($(Aa)!0.8!(Ab)$);
\draw[SA] (Ac) -- (Aa);
\draw[thick, dashed] (Ad) -- (Ac) node [midway, left] {{\textit{(E)}}};

\coordinate (Bb) at ($(B_QNAgent.north |- CC.south)$);
\coordinate (Bc) at ($(B_QNAgent.north)!0.4!(Bb)$);
\coordinate (Bd) at ($(B_QNAgent.north)!0.8!(Bb)$);
\draw[SA] (Bc) -- (B_QNAgent.north);
\draw[thick, dashed] (Bd) -- (Bc) node [midway, left] {\textit{(C)}};

\draw[SA] (CC_NCM.north) -- (CC_PGTC.south);

\end{tikzpicture}

 }
    \caption[Interaction Diagram for our proposed quantum network architecture.]{
        Interaction Diagram for our proposed quantum network architecture.
        Elements in red are local software components. 
        Single stroke arrows represent purely classical interactions and double-stroke arrows represent quantum interactions. 
        The dotted arrows denote the corresponding interaction from the other end node. 
        The \textit{QPS} and \textit{CPS} are the \textit{quantum processing system} and \textit{classical processing system} respectively. 
        The application stack includes the application code, compiler and execution environment. 
        The \textit{Network Capabilities Manager} is an oracle which can be queried by end nodes to find out information about the network as part of the network capabilities update phase of the architecture (\ref{ssec: Arch - NCU}).
        The process of \textit{Scheduler Admission} includes the demand queue. 
        The \textit{Quantum Network Stack} is that of \cite{dahlberg_link_2019}.
        The ellipsis represents the quantum network. 
        The labelled interactions are as follows: 
        \textbf{A}: Network Capability Update;
        \textbf{B}: Capability Negotiation;
        \textbf{C}: Demand Registration;
        \textbf{D}: Session Initialisation;
        \textbf{E}: Network Schedule Distribution (note this goes to all components of the network);
        \textbf{F}: Input of network schedule into the local schedule;
        \textbf{G}: Execution of the schedule(s). 
        }
    \label{fig:Network Interactions}
\end{figure*}
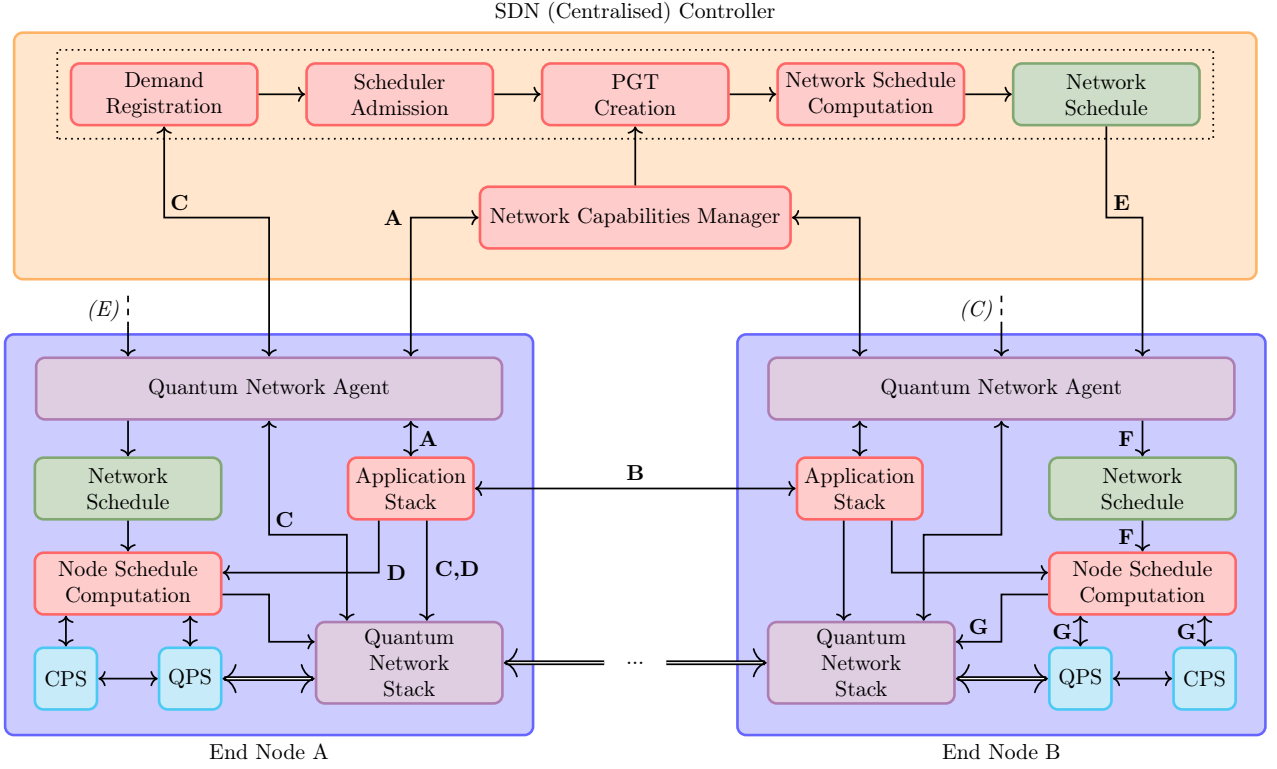

Following capability negotiation and demand submission, each end node performs some initial configuration (interaction \textbf{D} in Figure~\ref{fig:Network Interactions}). 
In particular, the blocks of instructions that make up the program are submitted to the local scheduler, and any initial configuration of the end node's quantum network stack is performed. 
An example of configuration of an end node's quantum network stack which may be required is the establishment of a quantum network socket for the application, as in \cite{vecht_qoala_2025, donne_design_2024}, ensuring that generated entanglement is assigned to the correct application. 

\subsection{Processing Demands and Packet Generation Tasks}\label{ssec: Arch - PGTs}

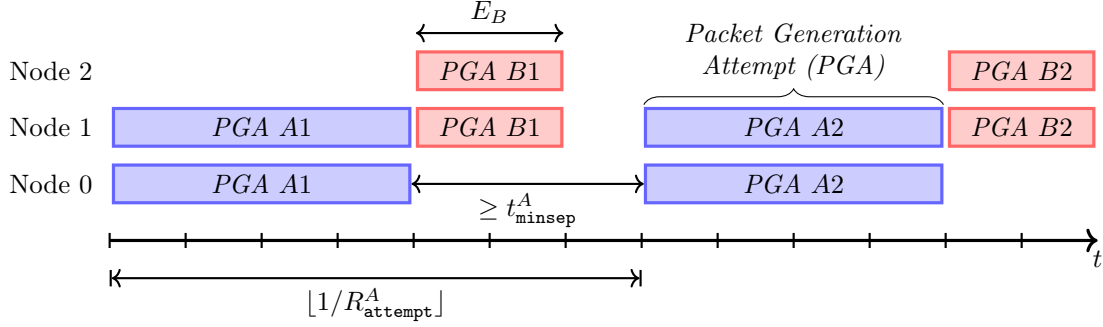
\begin{figure*}[t]
\centering
    \begin{tikzpicture}[node distance = 1cm,
            background rectangle/.style={fill=white},
            demandA/.style={rectangle,
                            draw=blue!60, 
                            fill=blue!20, 
                            very thick, 
                            align=center, 
                            minimum width=3.9cm, 
                            minimum height = 0.5cm,
                            },
            demandB/.style={rectangle,
                            draw=red!60, 
                            fill=red!20, 
                            very thick, 
                            align=center, 
                            minimum width=1.9cm, 
                            minimum height = 0.5cm,
                            },
            SA/.style={thick, ->,},
            DA/.style={thick, <->,},
            DAB/.style={thick, |<->|},
            show background rectangle,
			]

    \draw[SA, very thick] (0,0) -- (13,0) node [below] {$t$};
    \foreach \i in {0,...,12} {
        \draw[thick] (\i,0.1) -- (\i,-0.1);
    }

   \node[demandA] (A10) at (2,0.75) {\textit{PGA} $A1$};
   \node[demandA] (A20) at (9,0.75) {\textit{PGA} $A2$};

   \node[demandA] (A11) at (2,1.5) {\textit{PGA} $A1$};
   \node[demandA] (A21) at (9,1.5) {\textit{PGA} $A2$};

   \node[demandB] (B12) at (5,2.25) {\textit{PGA} $B1$};
   \node[demandB] (B22) at (12,2.25) {\textit{PGA} $B2$};

   \node[demandB] (B11) at (5,1.5) {\textit{PGA} $B1$};
   \node[demandB] (B21) at (12,1.5) {\textit{PGA} $B2$};

   \draw[DA] (A10.east) -- (A20.west) node [below, midway] {$\geq \tminsep^A$};

    \draw[DA] ($(B12.west |- 5,2.75)$) -- ($(B12.east |- 5,2.75)$) node [midway, above] {$E_B$};

    \draw[DAB] ($(0,0 |- 4,-0.5)$) -- ($(7,0 |- 4,-0.5)$) node [midway, below] {$\lfloor1/R_\texttt{attempt}^A\rfloor$};

    \node (label1) at ($(-0.1,0 |- A10.center)$) [left] {Node 0};
    \node (label2) at ($(-0.1,0 |- A11.center)$) [left] {Node 1};
    \node (label3) at ($(-0.1,0 |- B12.center)$) [left] {Node 2};

    \draw[decorate, decoration = {brace, amplitude=6pt}] ($(A21.west |- 2,1.8)$) -- ($(A21.east |- 2,1.8)$) node [midway, above, align=center, yshift=0.2cm] {\textit{Packet Generation}\\\textit{Attempt (PGA)}};

\end{tikzpicture}     \caption[Excerpt from an example network schedule with two PGTs]{Excerpt from an example network schedule with two PGTs $\tau_A = ((4,1/7,\{0,1,M\}),2,100)$ and $\tau_B = ((2,1/7,\{1,2,M\}),0,100)$ on three nodes, where $M$ represents a number of resources at a metropolitan hub. }
    \label{fig: example network schedule}
\end{figure*}

Whilst the demand format is sufficient for communicating the requirements of the nodes to the central controller, it is not an efficient format for use by a network scheduling algorithm. 
Therefore, once a demand has passed the demand registration phase, it is converted into a \textit{packet generation task} (PGT).
This is an internal representation of the demand containing only the information required to construct the network schedule.

\subsubsection{Finite Execution Times}

Due to the probabilistic nature of generating entangled links, it is impossible to guarantee that a packet will be generated when any finite exectution time is allocated to a \textit{packet generation attempt} (PGA). 
Therefore, the network only guarantees that a packet will be generated in a given PGA with some probability $p_\texttt{packet}$. This probability is an internal parameter, known only to the network. The rate at which PGAs are scheduled may be increased to compensate for $p_\texttt{packet}$.
The value of $p_\texttt{packet}$ can either be static, or can alternatively be determined using a method conforming to Algorithm~\ref{alg:determining p packet}.
Once the value of $p_\texttt{packet}$ has been determined, the exectution time of each PGA, $E$ can be determined, using a method conforming to Algorithm~\ref{alg:determing length of PGA}.

In the process of selecting $p_\texttt{packet}$, there is a trade-off between the length of each PGA, $E$, and the rate at which PGAs need to be scheduled, $R_\texttt{attempt}$.
In particular, as the value of $p_\texttt{packet}$ increases, the required length of the PGA increases, whilst the required rate of PGA decreases. 
Furthermore, as $E$ and $R_\texttt{attempt}$ change at different rates with $p_\texttt{packet}$, the resource utilisation $U_\tau := E_\tau R_\tau^\texttt{attempt}$ for a given packet generation task $\tau$ is not constant with $p_\texttt{packet}$. 
We will see in \S\ref{sssec: arx - NS - SAC} that the resource utilisation of packet generation tasks is an important quantity that determines how many or which demands can be accepted by the scheduler admission control.

\begin{algorithm}
    \SetKwInOut{Input}{Input}
    \SetKwInOut{Output}{Output}

    \Input{Network state, service agreements, $w$, $s$}
    
    \Output{Probability of generating a packet in a PGA, $p_\texttt{packet}$}

    Determine the value of $p_\texttt{packet}$ for given $w,s$ given network conditions and adhering to service agreements. 
    
    \caption{Determining $p_\texttt{packet}$}
    \label{alg:determining p packet}
\end{algorithm}

\begin{algorithm}
    \SetKwInOut{Input}{Input}
    \SetKwInOut{Output}{Output}

    \Input{$w$, $s$, $p_{succ}$, $p_\texttt{packet}$}
    \Output{Length of a PGA, $E$}
     Calculate the shortest time $E$ such that a packet $(w,s,F)$ is generated with probability $p_\texttt{packet}$ in time $E$, given a probability of entangled link generation $p_\texttt{succ}$.

    \caption{Determining the length of PGAs}
    \label{alg:determing length of PGA}
\end{algorithm}

\subsubsection{Setting the rate of packet generation attempts}

The next piece of information requiring computation by the central controller is a suitable rate, $R_\texttt{attempt}$, at which to schedule PGAs. 

If the nodes request a fixed rate, then the service model determines $R_\texttt{attempt}$. 
For example, if the service model is such that the requested rate is met exactly, then the central controller would set  $R_\texttt{attempt} = R/p_\texttt{packet}$.

If the requested rate is 0, that is the nodes are requesting the minimal rate to achieve minimal service, then the central controller needs to calculate this rate, using a method conforming to Algorithm~\ref{alg:determining minimum rate of packet generation}. 
In particular, the central controller sets a value $\epsilon_{service}$ for the maximum probability that any session does not obtain minimal service.
This parameter is known by both the central controller and the relevant end nodes, and is set by the network operator when initially setting the network up. 
From this, the minimum number of PGAs needed to meet the service threshold can be determined. Once this number is known, the minimum rate can be calculated.

\begin{algorithm}
    \SetKwInOut{Input}{Input}
    \SetKwInOut{Output}{Output}

    \Input{$\ninst, \texpiry$, current time, $\epsilon_{service}$, $p_\texttt{packet}$}
    \Output{Minimum possible rate at which to schedule PGAs.}
    Calculate the minimum number of PGAs, $N_{\texttt{min}}$, required such that the probability of at least $N_{\texttt{inst}}$ packets being generated is at least $1-\epsilon_{service}$\;
    \Return $N_{\texttt{min}}/(\texpiry - \texttt{current time})$

    \caption{Determining the minimum rate of PGAs}
    \label{alg:determining minimum rate of packet generation}
\end{algorithm}

If permitted by the service model and any agreements with nodes, the central controller also retains the right to reduce, or throttle, the rates requested. 
This may be done in order to reduce the load on the network due to either a particular demand or the collection of all demands.
Reducing the rate at which PGAs are scheduled may allow the network to serve more users at any one time, albeit at the cost of an increased probability that the throttled sessions do not obtain minimal service. 

\subsubsection{Additional Information}
The scheduler requires knowledge of the resource requirements $\rho$ of each demand. These depend on the path and entanglement generation protocol employed by the network to generate the desired entangled links, and can be obtained from the central controller's side of the network capabilities manager. In addition, the parameters governing the minimum separation time between subsequent PGAs, $t_{\texttt{minsep}}$, and the expiry time of the demand, $t_{\texttt{expiry}}$ are carried forward unchanged to the PGT. These final quantities are combined with the execution time of each PGA, $E$, and the rate at which PGAs should be scheduled, $R_{\texttt{attempt}}$, to define the complete PGT.

\subsubsection{Complete Packet Generation Task}
It is now possible to define the complete PGT. 

\begin{define}[Realisation of a demand]

      A \textit{realisation} of a demand is a specific choice of how PGAs from a packet generation task will be executed. 
      In particular, this includes specifying a protocol for generating end-to-end entangled links and which suitable packet will be generated.
      This also determines the path in the network along which links will be generated, and therefore the resources which will be required to execute each PGA.

      We write the set of all possible realisations as $\mathcal{R}$.
\end{define}

\begin{define}[Packet Generation Task]\label{define: PGT}
    A \textit{packet generation task} (PGT) is the following tuple:
    
    \begin{equation}
        \tau = \Big(\big\{(E,R_\texttt{attempt}, \rho)_{r}\big\}_{r\in \mathcal{R}},~ t_\texttt{minsep}, \texpiry\Big)\label{eq:PGT definition}
    \end{equation}
    where for each possible realisation $r\in\mathcal{R}$:
        \begin{itemize}
            \item $E$ is the execution time of each PGA
            \item $R_\texttt{attempt}$ is the average rate at which PGAs are scheduled.
            \item $\rho$ is the set of resources which are required to execute each PGA.
        \end{itemize}
\end{define}
The relation of these parameters to a network schedule can be seen in Figure~\ref{fig: example network schedule}.

We make the distinction between \textit{demands}, which pertain to the domain of the nodes and PGTs which pertain to the domain of the central controller in order to maintain separability in the control structure.
This separation reinforces that the central controller may determine how to realise a demand based on the information submitted and its knowledge about the state of the wider network.

If the network scheduling algorithm supports it, the packet generation task can be interpreted as a multi-mode task, similar to as in \cite{talbot_resource-constrained_1982}.
In such a situation, the network scheduler can choose the `best' mode in which to schedule the packet generation task given the current network load.
If such behaviour is not supported, then the central controller should reduce $\mathcal{R}$ to a single realisation/mode in which the PGT is always scheduled.

\begin{table*}[t]
    \centering
    \begin{tabular}{|c|c|}
        \hline
       Source  & Supplied Parameters \\\hline
       Application script &  $w,s,F, \tminsep$ \\
       End node(s) & $R, \texpiry, \ninst$ \\
       Central controller (configuration) & $p_\texttt{packet}, \varepsilon_{service}$ \\
       Central controller (calculation) & $E, R^\texttt{attempt}$\\
       Path and protocol for entanglement generation & $p_{succ}, \rho$ \\\hline
       
    \end{tabular}
    \caption[Table of entities (sources) and the  parameters they provide]{Table of entities (sources) and the parameters they provide.}
    \label{tab:my_label}
\end{table*}

\subsection{Network Scheduling}\label{ssec: Arch - NS}
Algorithm \ref{alg: network scheduling} summarises the procedure of network scheduling. The procedure is subdivided into the processes of scheduler admission control, computation of the schedule, and distribution of the schedule. Figure~\ref{fig:network scheduling timings} illustrates possible timings for each of these processes. 
\begin{algorithm}
    \SetKwInOut{Input}{Input}
    \SetKwInOut{Output}{Output}

    \Input{Demands in queue, previously scheduled PGTs, scheduling interval, service model}
    \Output{Network Schedule}
    Decide which, if any, demands to accept from the queue into the schedule\;
    Compute the schedule for the next scheduling interval following the chosen service model\;
    Distribute the schedule to all relevant parties\;
    
    \caption{Network Scheduling}
    \label{alg: network scheduling}
\end{algorithm}

\subsubsection{Scheduler Admission Control}

\label{sssec: arx - NS - SAC}
At the start of each scheduling interval,  the first task the central controller undertakes is to identify which PGTs should be given service in the next network schedule. 
Any PGTs from the previous schedule which have not expired or been terminated by the source nodes are automatically carried forward and new demands from the queue may also be admitted to the scheduler.

Any specific implementation of our architecture may define a tailored admission control routine which addresses the performance goals of the implementation. 
However, a guideline for any implementation is that the utilisation of any resource, $\mathcal U_r$, should satisfy $$\mathcal U_r = \sum_{\tau:r\in\rho_\tau}U_\tau \leq 1,$$ where $U_\tau = E_{r,\tau} R^\texttt{attempt}_\tau$ and $r$ is the specific realisation which is scheduled.
This is to ensure the schedules are feasible.

\subsubsection{Computing the Network Schedule}
Once the admitted PGTs are finalised, the central controller constructs the network schedule for the next scheduling interval, using a method conforming to Algorithm~\ref{alg: network scheduling}.
In doing so, each PGT which has been accepted for scheduling is assigned a series of start times, from each of which a PGA is executed without preemption.

Such a schedule can be time-slotted, though this is not required.
If a time-slotted schedule is employed, the central controller determines the length of the time slots, which are  the same for all resources in the network schedule.

The computed network schedule needs to cover the entirety of a scheduling interval. 
This can be achieved either by directly computing a schedule for the whole scheduling interval or by computing a shorter schedule that can be repeated to cover the whole scheduling interval. 

A constraint applies to the time required to compute the network schedule, and thus restricts the selection of a specific scheduling algorithm. This computation time should be short enough so that a single scheduling interval covers computation and distribution of the schedule, followed by a final buffer time for the construction of local schedules. End nodes require the final buffer to construct their local schedules, which incorporate the network schedule.

It is the role of an implementation to specify handling of the situation where the schedule is not computed in time.
An example of a policy which may be employed would be to delay execution of the late computed schedule to the start of the following scheduling interval. 
Alternatively, if the central controller distributes the schedule as soon as possible then each network component could start executing the schedule from the time when it is received.
Whichever policy is enforced, however, if there is a gap between the end of the previous schedule and the arrival of a  new schedule at a network component, a guideline for implementation is to instruct all components to only begin executing new PGAs upon the arrival of the new schedule. This ensures proper synchronisation across all components.

\subsection{Distributing the Network Schedule}
\label{subsec: Arch - distribution}
Once the schedule has been computed, it must be distributed to all components of the network.
Each component only receives the portion of the network schedule relevant to it. 
For example, an end node receives only its portion of the schedule, whereas a metropolitan hub receives the portion of the schedule concerning all nodes to which it is connected. 

 To ensure compatibility with requirements of the network stack and the runtime application execution environment on end nodes, the format of the schedule must be such that the relevant end nodes receive the start and end times of each scheduled PGA together with an identifier of which demand the scheduled PGA pertains to. Various implementations are possible.

As indicated in Figure~\ref{fig:network scheduling timings}, sufficient time must be allotted for distribution of the network network schedule. The interval reserved for distribution should be such that the probability that the any component does not receive the network schedule is vanishingly small. 
However, in the case that a component does not receive the schedule on time, we expect that the affected components will continue to request the network schedule from the central controller.
Furthermore, as with the case where the schedule is not computed on time, such components should not start any new PGAs until they receive the correct schedule. 

Once a network schedule is received by an end node, it can compute its local schedule.
This uses the network schedule to determine when submitted QC (entanglement generation) blocks are executed (interaction \textbf{F} in Figure~\ref{fig:Network Interactions}). The remaining instruction blocks are then scheduled relative to these QC blocks \cite{vecht_qoala_2025, jirovska_evaluating_2023}. 

Once this process has been completed, the local and network schedules can be executed, without any extra interaction either between pairs of end nodes and between end nodes and any internal network components (interaction \textbf{G} in Figure~\ref{fig:Network Interactions}).

\section{Implementation}\label{sec: Example Implementation}

In order to be able to perform an evaluation in \S\ref{sec: Evaluation}, we provide an example implementation of the central controller in our architecture, focusing on the process of computing the network schedule.  
All the algorithms explicitly presented in this section are applicable to any network topology, with any applications being executed over the network.
Therefore, this implementation can act as a baseline against which future implementations can be compared.

\subsection{Service Model}
\label{subsec:Imp - Service Models}

 For our implementation, the network operates under the following network model:

\begin{nsmodel}[No Throttling]
Under this model, demands are met exactly, that is PGAs are scheduled to on average meet exactly the rate requested by the nodes. 
In the case where the network is oversubscribed, then demands which cannot be accepted to be scheduled are delayed until there is space, or dropped if they can never be served.
\label{sm: No Throttling}
\end{nsmodel}
Note that this service model requires some admission control for the scheduler, to decide which demands are accepted for scheduling and which are delayed. 
We give some examples of such rules in \S\ref{subsubsec: imp - NS - admin control}. 

This service model is particularly applicable to scenarios where nodes are relying on demands being met exactly.
An example of such a scenario would be using QKD to underpin the availability and monitoring of secure critical infrastructure. 
In such a case, a demand being throttled could potentially result in loss of access to the infrastructure, due to not being able to generate keys quick enough. 
Examples of such schemes include \cite{green_quantum_2023, alshowkan_authentication_2022, aggarwal_authentication_2024}.

\subsection{Network Capability Update}
The evaluation will be performed on a network with a star topology (Figure~\ref{fig:implementation star network}), as it removes any influence from routing or determining schemes for generating end-to-end entangled links.
Consequently, we will not give implementations of such schemes.  
For examples of how routing may be performed in more complicated network topologies, see for example \cite{pant_routing_2019} and \cite{van_meter_path_2013}.

\subsection{Capability Negotiation}
We assume that capability negotiation is carried out using the \textit{exposed hardware interface (EHI)} from Qoala \cite{vecht_qoala_2025}.
This allows the end nodes to exchange information about the hardware and software constraints of their devices. 

\subsection{Demand Registration and Queuing}
We use the following set of rules for demand registration: 
\begin{drrule}
    The demand must be sane, in particular, $w\geq s$ and both parties are capable of generating entangled links. 
\end{drrule}
\begin{drrule}
    Let the utilisation of a PGT $\tau$ with execution time $E$ and period $T$ be $U_\tau=E/T$.
    Then the utilisation of a PGT resulting from at least one of the requested service options must be less than $\hat U = 0.8$. 
    \label{dr rule: utilisaiton}
\end{drrule}

Once demands are registered, they are placed into a \textit{first-in-first-out} (FIFO) queue for consideration by the scheduler admission control.

\subsection{Creating Packet Generation Tasks}
We use a fixed value for the probability, $p_\texttt{packet}$, that a packet is successfully generated by a PGA.
Then, to determine the length of a PGA, we use results from \textit{scan statistics}.
Specifically we use the approximations in \cite{naus_approximations_1982} (also given in Appendix~\ref{app: Naus Approximation}) for the probability of $k$ successes in a window $w$ given a total of $N$ trials. 
From this we can calculate the minimum execution time of the PGA, such that the PGA succeeds with probability $p_\texttt{packet}$, using interval bisection.

If the nodes have requested the central controller calculate the minimum rate ($R=0$), we use Hoeffding's inequality to calculate the minimal $N_{min}$ such that after $N_{min}$ PGAs, the session obtains minimal service with probability $1-\epsilon_{service}$. 
The rate of packet generation attempts is then set by ${N_{min}}/{(\texpiry - \texttt{[current time]})}$.
This process is described in detail in Appendix~\ref{app: PGT creation - Min Rate Hoeffding}.
Otherwise $(R>0)$, we simply set $R_\texttt{attempt} = p_\texttt{packet}^{-1}R$

\subsection{Network Scheduling}
\label{subsec: implimentation - Network Scheduling}
\subsubsection{Admission of New Demands}
\label{subsubsec: imp - NS - admin control}
The central controller decides which demands to remove from the queue by checking if accepting a demand would violate either of the following rules:
\begin{acrule}[Utilisation Bound]
    For all resources $r$, the utilisation of $r$, $\mathcal U_r$, must satisfy
    \begin{equation}
    \mathcal U_r = \sum_{\tau:r\in\rho_\tau}U_\tau \leq \resourceutilisationbound
\end{equation}
    for some constant $\resourceutilisationbound\in (0,1]$.
    \label{sac rule:utilization} 
\end{acrule}
\begin{acrule}[Computation Time Bound]
    The estimated time to compute the network schedule cannot exceed $\alpha_C T_{SI}$, for some constant $\alpha_C\in(0,1)$.
    \label{sac rule:computationBound}
\end{acrule}

If neither rule is violated, then the demand is accepted and the corresponding PGT created. 
Otherwise, unless \ref{dr rule: utilisaiton} is violated, the demand is returned to the head of the queue.
Note that as we use a \textit{FIFO} queue demands can only be accepted in the order they were submitted, so if a given demand fails admission control, no more demands can be accepted from the demand queue. 

 $\resourceutilisationbound$ is restricted to $(0,1]$ as if any $\mathcal{U}_r > 1$, then there is not physically enough time to schedule all the required PGAs. 

Note as well that the estimated computation time in \ref{sac rule:computationBound} may be continuously updated by the central controller based on its current performance when computing network schedules. 
Furthermore, we cannot take $\alpha_C\geq1$ as there must be some time remaining in the scheduling interval to distribute the schedule, as per Figure~\ref{fig:network scheduling timings}.

We write $\mathcal T$ to be the set of PGTs from demands which have passed admission control and have not expired or been terminated. 
Once a demand has been terminated or expires, then no more PGAs will be scheduled from it. 
Therefore $\mathcal T$ is the set of PGTs from which we need to schedule PGAs in the next network schedule. 

\subsubsection{Computing the Network Schedule}

To calculate the network schedule, we adapt priority-based periodic task scheduling methods from real time systems, and use a time-slotted network schedule, similar to as in \cite{skrzypczyk_architecture_2021}. 
In particular, we will use \textit{earliest deadline first (EDF)} scheduling, and adapt the following model, which is common to real-time periodic scheduling (\textit{c.f.}, \cite{stankovic_deadline_1999, liu_scheduling_1973}):
\emph{A periodic task creates jobs.
Each of these jobs is released at some time $r$, only after which can it be executed, and must be completed by a deadline $d$.
If a job is the $i$th released by a task, then its release time is given by $r=(i-1)T+\sigma$ where $T$ is the period of the task and $\sigma$ is an offset determining how long after the start of the schedule the first job is released. 
The corresponding deadline is then set to the start of the next period, \textit{i.e.} $d=r+T=iT+\sigma$.}

In our case, the tasks are PGTs and the jobs are PGAs. 
Following the definitions in \cite{stankovic_deadline_1999}, we take our deadlines to be \textit{soft}, which permits PGAs to be scheduled past their deadlines if it is not possible to do otherwise.
In this way, no PGAs will be skipped, and so the average rate of packet generation experienced by the end nodes will be as they requested over the course of the entire lifetime of the demand. 

To be able to determine the release times and deadlines of each PGA, the period of the PGT is required.
This is given by 
\begin{equation}    
T_{\texttt{attempt}} = \left[\frac{1}{R_\texttt{attempt}t_{timeslot}}\right]_\mathbb{N}, \label{eq: period of PGT}
\end{equation}
where we write $[x]_\mathbb{N}$ to be $x$ rounded to the nearest positive integer and $t_{timeslot}$ is the duration of a timeslot in the network schedule.
We also need to determine the offset $\sigma_i$.
This is set to be the start of the scheduling interval during which PGT $\tau_i$ is first scheduled. 

We can now determine the release time and deadlines of each PGA which needs to be scheduled.
For $\tau_i\in\mathcal T$, let $\tau_{i,j}$ be the $j$th PGA from PGT $i$. 
Let $T_i$ and $\sigma_i$ be the period and offset of $\tau_i$ respectively, let $r_{i,j}$, $d_{i,j}$ be the release time and deadline of $\tau_{i,j}$, and let  $s_{i,j}$ and $c_{i,j}$ be the start and completion times of $\tau_{i,j}$, respectively.
Then, \begin{align}
    r_{i,j} &= \sigma_i + \max\{(j-1)T_i,c_{i,j-1} + t_\texttt{minsep,i}\}\label{eq:PGA release times}\\
    d_{i,j} &= \sigma_i + jT_i\label{eq: PGA deadlines}
\end{align}
where $j=1,...,\left\lfloor{(t_{\texttt{expiry}, i} - \sigma_i})/{T_i}\right\rfloor$.
Note that we adapt the determination of the release times from the usual formula in order to incorporate the minimum separation time between two PGAs.

\begin{define}[Eligibility for scheduling of a PGA]
    We say that a PGA $\tau_{i,j}$ is \textit{eligible} for scheduling at time $t^*$ if it meets the following criteria:
    \begin{enumerate}
        \item It has been released ($t^* > r_{i,j}$)
        \item It has not yet been scheduled (the start time $s_{i,j}$ is undefined)
        \item The required resources are available
    \end{enumerate}
\end{define}

The schedule is then computed according to Algorithm~\ref{alg:priority-scheduling-algorithm}.

\begin{algorithm}
    \SetKwInOut{Input}{Input}
    \SetKwInOut{Output}{Output}

    \Input{Desired schedule length, PGTs to be scheduled, $\mathcal T$}
    \Output{Scheduled PGAs}

    Set the initial decision time $t^*$ to the time at the start of the schedule\;
    \While{$t^*$ is less than the end time of the schedule}{
        \While{there are eligible PGAs}{
            Schedule the eligible PGA with the earliest deadline\;
            Update the set of eligible PGAs\;
        }
        Update $t^*$ to the next time either a task is released, a task completes execution, or the end of the scheduling interval, whichever is earlier \;
    }
    \caption{EDF Scheduling Algorithm. \label{alg:priority-scheduling-algorithm}}
\end{algorithm}

\subsection{End Node Scheduling}

As our evaluation focuses on the network portion of the architecture, we do not compute the local schedules. 
We assume that node schedules exist which facilitate the successful execution of an application instance given that packet generation succeeds.
For examples of methods of constructing schedules on end nodes, see for example the scheduler built into the Qoala environment simulation in \cite{vecht_qoala_2025} or one of the schedulers used in \cite{jirovska_evaluating_2023, QNodeOS}.

\section{Evaluation}\label{sec: Evaluation}

\begin{figure}
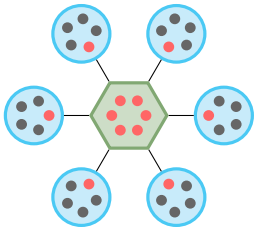

    \centering
    \starNetwork{6}{5}
    \caption[Example of a star-topology network with 6 nodes.]{Example of a star-topology network with 6 nodes. The outer circles represent the end nodes and the central hexagon a central junction node. In each network time-slot the central node can attempt to create an entangled link with each of the end nodes, and then perform entanglement swaps to create end-to-end links between pairs of end nodes. The orange dots represent communication qubits and the black dots represent memory qubits. }
    \label{fig:implementation star network}
\end{figure}

We now evaluate the performance our network architecture using the implementation we described in the previous section.
For the simulations we perform, we use two test applications from different application classes. 
We define the performance metrics we consider, and where relevant we provide further details about the precise model used in our simulations.
Further details about the simulation model can be found in Appendix~\ref{app:Additional Eval Details}.
From the results we obtain we  validate the viability of our architecture and are able to draw conclusions about the need for good admission control and how nodes should decide what rates of packet generation to request.

\subsection{Metrics}
\label{subsec:eval - Metrics}
\begin{define}[Termination of a Session]
    We say a session is \textit{terminated} if the session has obtained minimal service and sent a termination message to the central controller.
    This results in no further time being allocated for this demand beyond the end of the schedule currently being computed. 
\end{define}
\begin{define}[Expiry of a session]
    We say a session is \textit{expired} if the expiry time has elapsed without a termination message being received.
    No packet generation attempts will be scheduled after the expiry time. 
\end{define}
Note that a session can expire and still have obtained minimal service, if the nodes choose not to send the termination message and carry on generating packets.
Likewise, a node could choose to send a termination message before obtaining minimal service. 
However, we assume in our evaluation that neither of these behaviours occur. 

\begin{metric}[Proportion of Expired or Terminated Sessions Obtaining Minimal Service]
Let $\mathbb{S}$ be the set of application sessions initiated in a given simulation. 
Let $\hat{\mathbb{S}}\subseteq\mathbb{S}$ be the set of those application sessions which are expired or terminated.
Let $\mathcal{M}\subseteq\hat{\mathbb{S}}$ be the set of those application sessions which obtained minimal service.
Then the metric, $p^{MS}$, is given by 
\begin{equation}
\label{eq:propMinimalService}
p^{MS} = \frac{|\mathcal{M}|}{|\hat{\mathbb{S}}|} \in [0,1].
\end{equation}
    
\end{metric}

Such a metric is important to both the end nodes and the central controller. 
End nodes can use this to estimate the likelihood of a submitted application session obtaining minimal service, and therefore what quality of service the network can provide. 
On the other hand, by monitoring this metric, the central controller can assess the effectiveness of the (scheduler) admission control rules given the traffic on the network, and update them accordingly.

\begin{metric}[Average Time Spent in Queue]
Let $t_{submit,\mathcal S}$ be the time the demand corresponding to session $\mathcal S$ is submitted to the central controller. 
Let $t_{exit,\mathcal{S}}$ be the time that the demand corresponding to session $\mathcal S$ leaves the queue. 
Then the metric is given by
\begin{equation}
  \bar{t}_{queue} = \frac{1}{|\mathbb S|}\sum_{\mathcal S \in \mathbb S}\bigg(t_{exit,\mathcal S} - t_{submit,\mathcal S}\bigg).
\end{equation}
\end{metric}

For end nodes, evaluating this metric gives them an estimation of the expected latency between submitting a demand and receiving service. 
This in turn can then be used to estimate the load on the network, and even inform the choice of expiry time for the application session.
For the central controller, this metric gives an estimation of how overloaded the network is, especially in conjunction with $p^{MS}$.
Using this, the central controller can control the overload by either adapting the admission control rules as required, or even requesting that nodes reduce the rate at which new demands are submitted. 

\subsection{Network Configuration}

We consider a network consisting of 6 end nodes connected to a single central node as shown in Figure~\ref{fig:implementation star network}. 
Each end node is equipped with 5 qubits, of which at most one can be used for entanglement generation at any given time. 
The central node also has 6 independent communication qubits with no storage capability beyond the current network time slot.
We assume that in every time slot, an elementary entangled link may be generated between each end node and the central node.

Links with the central hub are consumed by performing deterministic entanglement swapping operations to generate end-to-end entangled links between pairs of end nodes.
The swap operations which are carried out are determined by which pairs of end nodes have a PGA scheduled in that time slot. 
We assume that the central node has no quantum memory and therefore any link which is generated must be consumed within the same time slot or be lost. 

We therefore simulate a mathematically equivalent model, whereby for each disjoint pair of nodes, an end-to-end link may be generated in each time slot with fixed probability ${p}_\text{gen}\ll1$.

\begin{table}[t]
    \centering
    \begin{tabular}{|c|c|}
       \hline
         Parameter &  Value \\\hline
         Network time-slot length& 100\,$\mu$s \\
         $p_{gen}$ & $7.5\times10^{-5}$\\
         $F$ & 0.925\\
         qubits at end node & 5\\\hline
    \end{tabular}
    \caption[Network Parameters for Evaluation]{Network Parameters for Evaluation. $p_{gen}$ is the probability of an end-to-end entangled link being created in a given time-slot. The trends we observe are generally insensitive to the value of $p_{gen}$.
}
    \label{tab:network paramters main}
\end{table}
\begin{table}[t]
    \centering
    \begin{tabular}{|c|c|}
       \hline
        Parameter &  Value \\\hline
        $p_\texttt{packet}$ & 0.2\\
        $\epsilon_{service}$ & $10^{-5}$\\
        PGA cap per schedule & 1500\\
        Utilisation Bound per link, $\resourceutilisationbound$ & 0.85\\
        Computation time factor, $\alpha_C$ & 0.5\\
        Scheduling Interval (\verb|MDA||\verb|CKA|) & 300s | 3600s\\\hline
    \end{tabular}
    \caption[Scheduler Parameters for Evaluation]{Scheduler Parameters for Evaluation. Except where noted, the trends that we observe are insensitive to the specific values chosen here.}
    \label{tab:scheduler paramters main}
\end{table}

\subsection{Admission Control}

\subsubsection{Utilisation Bound}

We set the value of the utilisation bound $\resourceutilisationbound = 0.85$. 
A restriction to $\resourceutilisationbound\leq1$ is necessary to ensure the schedule is feasible. 
However, as the tasks are non-preemptable, this is not a sufficient condition because of so-called priority inversions.
These occur when a task is available to be scheduled, but there is a non-preemptable lower priority task which is currently being executed preventing the higher priority task from being scheduled \cite{liu_scheduling_1973}.
As it is non-trivial to determine if a deadline will be missed without computing the schedule \cite{HardRealTimeComputing}, we instead reduce the utilisation bound to reduce the probability that a priority inversion causes a missed deadline.
By performing additional simulations beyond those reported on here, we observed that the conclusions we draw hold for any value of $\hat{\mathcal{U}}<1$.

\subsubsection{Schedule Computation Time}
We set $\alpha_C=0.5$, i.e. network schedules cannot take more than half the scheduling interval to compute. 
However, to avoid our results depending on the hardware of the server on which we perform our simulations, we do not directly implement \ref{sac rule:computationBound}. 

As the time to compute the schedule depends on the number of PGAs which will need to be scheduled,  the central controller can implement this rule by imposing a cap on the estimated number of PGAs which need scheduling in a given scheduling period. 
Therefore, we simply fix this cap \textit{a priori} and do not update it based on the observed computation times. 

The central controller will be running on dedicated hardware in any actual deployment.
This means that we do not expect the computation times to fluctuate much as there are no other background processes consuming CPU resources. 
Therefore, despite fixing this cap on the number of PGAs per schedule to be constant, the results obtained will still be indicative of real-world performance. 

In our simulations, we set this cap to 1500 PGAs per schedule. 
This is motivated by empirical characterisation of the time our computation server takes to calculate the network schedule for various number of PGAs, more details of which can be found in Appendix~\ref{app: eval supp - PGA cap}.

\subsection{Session Model}
\label{ssec: eval - session model}
\subsubsection{Creation time}

Let $t_\mathcal{S}^{MS/E}$ be the time at the end of the scheduling interval during which application session $\mathcal S$ either expires or obtains minimal service, and let $t_{renew}$ be an exponentially distributed waiting time with parameter $\lambda$.
Then the nodes involved in $\mathcal S$ begin a new session $\mathcal S'$ of the same application at time $t_{\mathcal S}^{MS/E} + t_{renew}$.
 In each simulation, the value of $\lambda$ is the same for all pairs of nodes.

\subsubsection{Contents}
In each of our simulations, all pairs of nodes execute the same application sessions.
Two test applications are considered: the first is a measure-directly application motivated by QKD (hereafter \verb|MDA|), and the second is a create-and-keep application motivated by the blind quantum computing algorithm in \cite{leichtle_verifying_2021} (hereafter \verb|CKA|). 
 For further discussion of the distinction between measure-directly and create-and-keep application classes refer to Section \ref{subsec:DesignConsiderations:ApplicationClasses}.
 The Qoala files we use for these test applications are given in Appendix~\ref{app: eval - qoala files}.

We fill out the remaining fields in the application sessions as follows:
\begin{equation}
    \mathcal{S} = (\mathcal{N}, \texttt{APP},\ninst=100, \texpiry=t_{submit} + t_\texttt{max duration})
\end{equation}
where $\texttt{APP} \in \{\texttt{MDA}, \texttt{CKA}\}$, $t_{submit}$ is the time the corresponding demand is submitted to the central controller and $t_\texttt{max duration}$ is the maximum duration of a session.

We acknowledge that the value of $\ninst=100$ is very low, as for example a typical QKD application would require $\ninst\gg10^5$.
However, we make this choice for simulation purposes, as it cuts down the required run time of our simulations (some of which run in almost real-time).
We expect that the conclusions drawn from our simulation results remain valid when $\ninst$ is increased to values which more accurately reflects the requirements of real-life applications.

\subsubsection{Peer-to-Peer vs Client-Server}
We consider two regimes under which pairs of nodes execute sessions, \textit{peer-to-peer} and \textit{client-server}. 
In the peer-to-peer regime, an end node can undertake an application session with any other end node. 
This is the type of behaviour we would expect for applications such as QKD, where every end node in the network is capable of carrying out the application.

In the client-server regime, one of the nodes is designated as the `server', and the rest as `clients'.
Client nodes are only able to undertake sessions with the server and not with each other. 
This type of behaviour we would expect for applications such as BQC, where one end node needs to be much more powerful than the other. 

We use the first regime for modelling \verb|MDA| traffic and the second for \verb|CKA| traffic.

\subsubsection{Termination of Demands}

We assume that nodes will send a message to the central controller to terminate their demand as soon as they obtain minimal service. 
Once a demand is terminated, it is not considered further for scheduling. 
Nodes will, however, continue to execute application instances whilst there is time allocated in the network schedule.

\subsection{Results}
\begin{figure*}[t]
    \centering
    \begin{subfigure}[b]{0.48\textwidth}
        \centering
\resizebox{\linewidth}{!}{\includegraphics{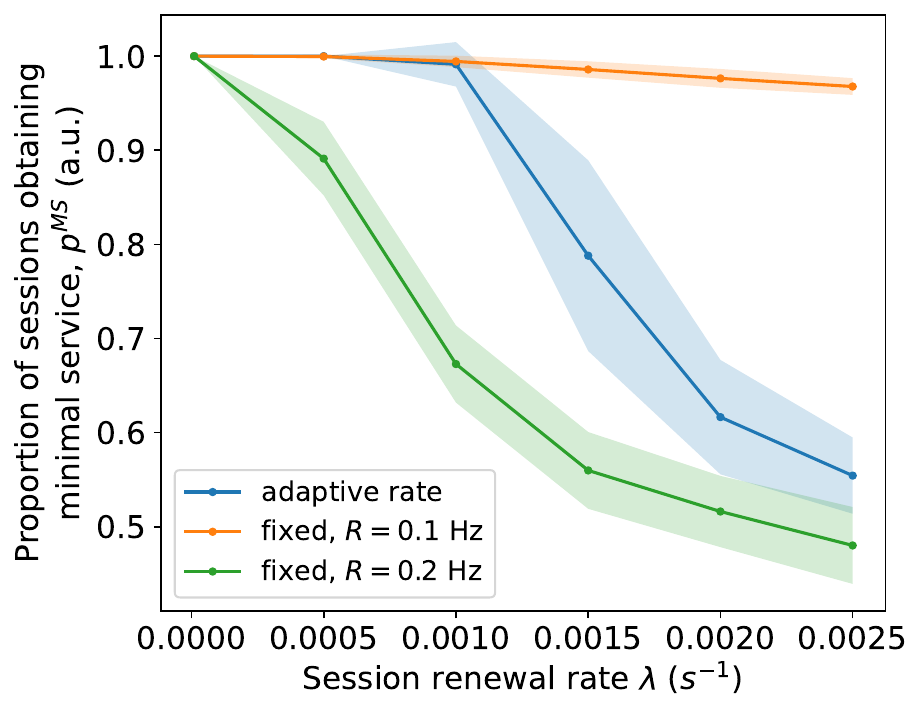}}
        \caption{}
        \label{fig:eval-results-p2p_mda-prop_min_service}
    \end{subfigure}
    \begin{subfigure}[b]{0.48\textwidth}
        \centering
\resizebox{\linewidth}{!}{\includegraphics{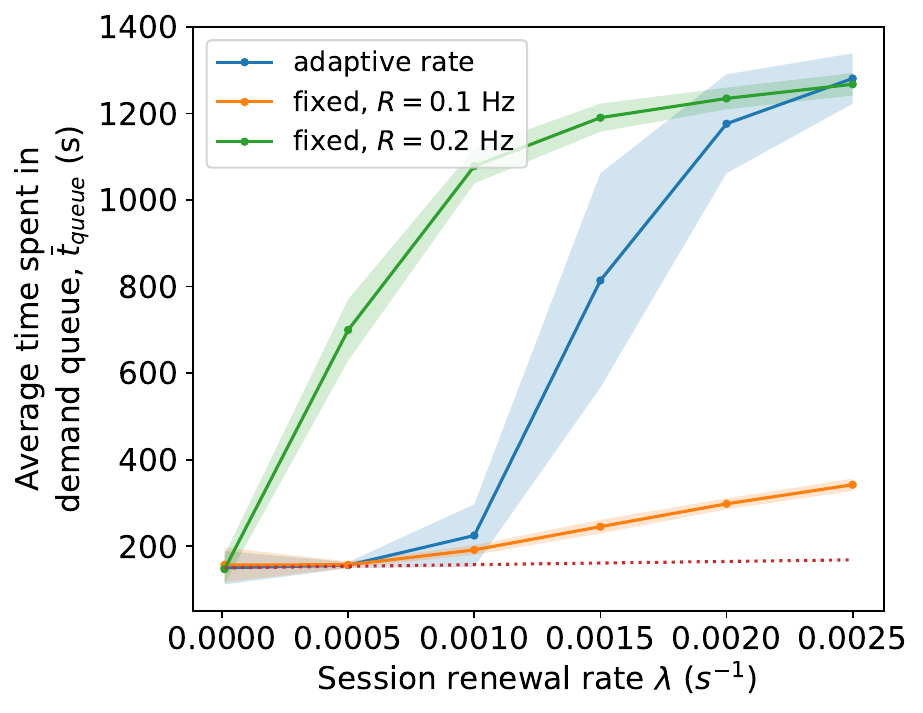}}
        \caption{}
        \label{fig:eval-results-p2p_mda-latency}
    \end{subfigure}
    \caption[Results from simulations for peer-to-peer \texttt{MDA} on a six-node star network.]{Results from simulations for peer-to-peer \texttt{MDA} on a six-node star network.
    (a) Proportion of initiated sessions which obtained minimal service from the network.
    (b) The average time a demand spent in the demand queue. 
    The shaded area represents $\pm 1\sigma$.
    The red dotted line represents the expected value of $\bar t_{queue}$ for a single pair of end nodes submitting demands. 
    The total simulated time was 6 hours.
    To produce each data point we average the results of 100 simulations, with simulations where zero sessions were initiated removed. There were no such simulation runs observed.}
    \label{fig:eval-results-MDA}
    \end{figure*}

\begin{figure*}[t]
    \centering
    \begin{subfigure}[b]{0.48\textwidth}
        \centering
\resizebox{\linewidth}{!}{\includegraphics[]{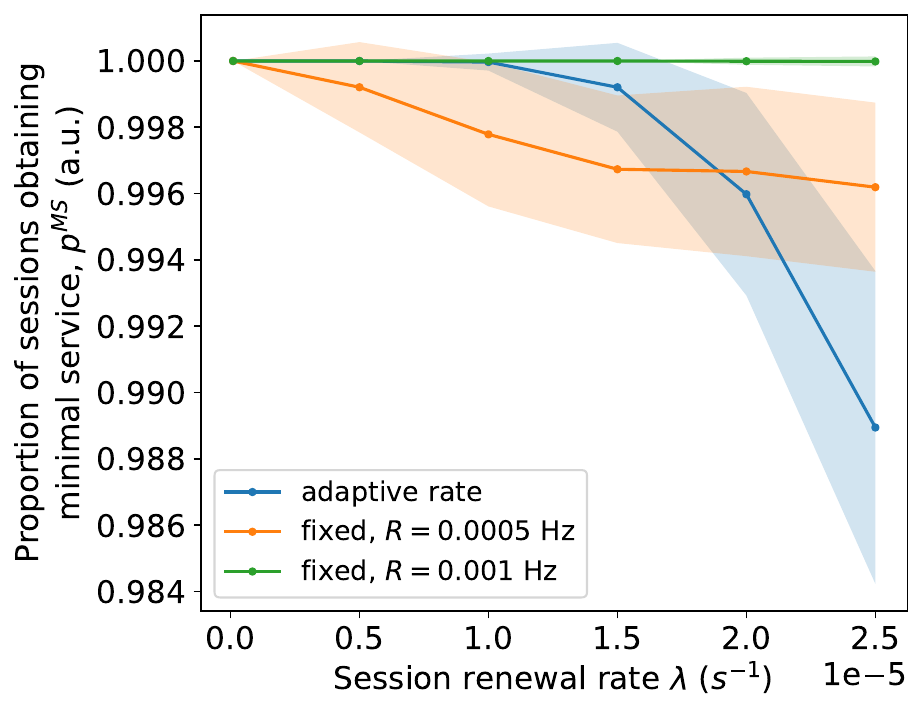}}
        \caption{}
        \label{fig:eval-results-cs_bqc-prop_min_service}
    \end{subfigure}
    \begin{subfigure}[b]{0.48\textwidth}
        \centering
\resizebox{\linewidth}{!}{\includegraphics[]{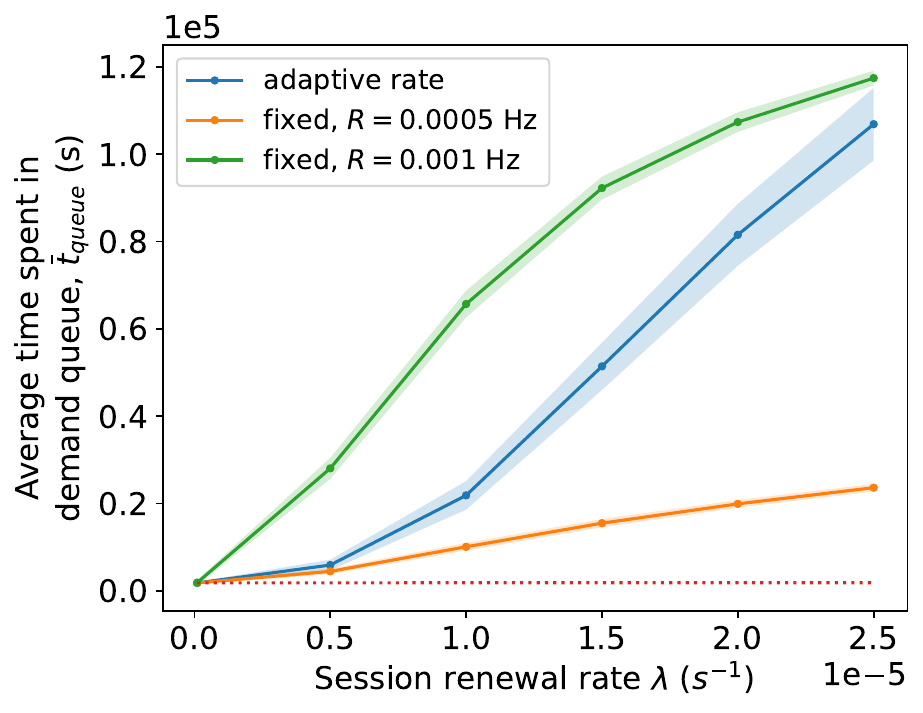}}
        \caption{}
        \label{fig:eval-results-cs_bqc-latency}
    \end{subfigure}
        \caption[Results from simulations for client-server \texttt{CKA} on a six-node star network.]{Results from simulations for client-server \texttt{CKA} on a six-node star network.
    (a) Proportion of initiated sessions which obtained minimal service from the network.
    (b) The average time a demand spent in the demand queue.
    The shaded region represents $\pm 1 \sigma$. 
    The red dotted line represents the expected value of $\bar t_{queue}$ for a single pair of end nodes submitting demands. 
    The total simulated time was 360 days. 
    All data points are the average of 250 simulations, with simulations where zero sessions were initiated removed. There were no such simulation runs observed.}
    \label{fig:eval-results-BQC}
\end{figure*}

Simulation of the implementation (Section \ref{sec: Example Implementation}) of our network architecture is an opportunity to confirm the feasibility of our architecture and to asses the performance of an implementation, as quantified by the metrics in Section \ref{subsec:eval - Metrics}.
These metrics may depend on parameters that are specific to how users demand service from the network as well as on the types of applications from which user demands originate.
We focus on assessing how changes to the parameters $\lambda$ and $R$, respectively the session renewal rate and the requested rate of packet generation, impact the values obtained for the performance metrics.
To asses the impact of the type of application, we also compare the results of simulations where all nodes are running the \verb|MDA| test application in the peer-to-peer regime (Figure~\ref{fig:eval-results-MDA}) to simulations where all nodes are running the \verb|CKA| application in the client-server regime (Figure~\ref{fig:eval-results-BQC}).


\subsubsection{Facilitation of obtaining minimal service}

 The most important criteria that a network architecture must meet in order to be considered viable is the capability to meet user demands. Here we quantify the degree to which an implementation of our architecture meets user demands with the performance metric \textit{proportion of expired or terminated sessions which obtain minimal service}, $p^{MS}$, quantified by (\ref{eq:propMinimalService}).
For each of the test applications simulated we observe that the implementation of our network architecture is able to successfully deliver minimal service to some proportion of sessions.
For the \verb|MDA| test application operated in the peer-to-peer regime (Figure \ref{fig:eval-results-p2p_mda-prop_min_service}) the proportion of demands which obtain minimal service is always at least $0.45$, for all parameter combinations simulated.
In complement, we observe that for the \verb|CKA| test application operated in the client-server regime a much higher proportion (>0.98) of sessions obtain minimal service, for each combination of the parameters simulated.

This general increase in the value of $p^{MS}$ between the client-server \verb|CKA| and peer-to-peer \verb|MDA| simulations can be attributed in part to the decrease in the number of possible demands. 
In the client server regime, there are only 5 sources of demands, whereas in the peer-to-peer regime there are 15 possible sources of demands. 
As each source only has a single demand registered at any time, the maximum length of the queue is much shorter in the client-server \verb|CKA| simulations.
This in turn means that as a fraction of the value of $\tmaxdur$, the average queuing times are much shorter in the \verb|CKA| simulations, which leads to greater values of $p^{MS}$.

From Figures \ref{fig:eval-results-p2p_mda-prop_min_service} and \ref{fig:eval-results-cs_bqc-prop_min_service} it is clear that the session renewal rate $\lambda$ and the requested rate of packet generation $R$ have a large impact on the proportion of sessions which obtain minimal service.
In particular, as $\lambda$ increases, the proportion of sessions which obtain minimal service may only decrease.
This is the expected behaviour, as an increase in $\lambda$ directly translates into increased load on the network.
We do note that in the case where $R=0.001$Hz in Figure~\ref{fig:eval-results-cs_bqc-prop_min_service}, $p^{MS}=1$ for all $\lambda$, with only a small non-zero standard deviation for $\lambda > 2\times10^{-5}$.
This indicates that there are parameter regimes where it is possible for all nodes to obtain minimal service, regardless of the session renewal rate. 
Figures \ref{fig:eval-results-p2p_mda-latency} and \ref{fig:eval-results-cs_bqc-latency} confirm that the average time a demand spends in the queue, $\overline{t}_{queue}$ always increases as $\lambda$ increases.
An extension of the duration of time that demands spend queued is direct evidence of increased load on the network.

The effect of $R$ is more subtle, as we consider two values of a fixed rate as well as an adaptive rate.
In the case of the \verb|MDA| test application operated in peer-to-peer mode (Figure \ref{fig:eval-results-MDA}) the lower fixed rate results in higher $p^{MS}$ for all values of $\lambda$.
In contrast, for the \verb|CKA| test application operated in client-server mode, the lower fixed rate results in lower $p^{MS}$.
See \S~\ref{sssec:eval - results - fixed rate comparison} for an explanation for why this is the case. 
However, as could be expected, for both test-applications the lower fixed rate results in lower $\bar{t}_{queue}$. This is expected because a lower fixed rate again translates to lower load on the network.

In section \S\ref{subsubsec: eval - results - adaptive}, we comment on the effectiveness of adaptive rate requests, in comparison to requests for fixed rates.

\subsubsection{Admission Control Requirements}

Whilst being able to facilitate some application sessions obtaining minimal service is sufficient for our architecture to be viable, it is desirable that this is the case for as many sessions as possible. 
In the peer-to-peer \verb|MDA| simulations, however, we observe that the value of $p^{MS}$ drops below 0.6 for the adaptive rate demands and below 0.5 for the fixed $R=0.2$Hz requests as $\lambda$ increases. 
This is due to the increased load which these demands place on the network, compared to the demands with a fixed rate of $R=0.1$Hz. 
In particular, as each demand/PGT requires a greater utilisation of the links to the central junction node, fewer demands can be served at any one time.
This leads to increased queuing times as observed in Figure~\ref{fig:eval-results-p2p_mda-latency}.
Therefore, a greater number of demands reach their expiry time without getting scheduled, and moreover those demands which are accepted for scheduling have much less time for generating packets, leading to a lower probability of obtaining minimal service. 

To prevent the value of $p^{MS}$ from dropping so far, the central controller should not allow the network to become so overloaded. 
This may be achieved through the admission control rules employed, both at demand registration and scheduler admission. 
In particular, we expect that if a demand is rejected, for example for requesting too high of a rate of packet generation, then the nodes would re-submit this demand with, for example, lower values of $R$.
From the data we observed, we can conclude that the demand registration rules in our implementation were not stringent enough for the peer-to-peer \verb|MDA| scenario. 
In particular, the fixed rate $R=0.2$Hz demands probably should almost always have been rejected, as by accepting them it led to a situation where a significant proportion of sessions were not able to obtain minimal service from the network (up to 50\% by $\lambda=0.0025$).
However, the same set of admission control rules was sufficient for the client-server \verb|CKA| scenario ($p^{MS}>0.98$ for all $\lambda$). 
Therefore, when designing the set of admission control rules to implement, not only should the properties of the demands themselves be taken into account, but also the frequency and number of demands which the central controller expects to receive.

\subsubsection{Controls on Requesting Adaptive Rates}
\label{subsubsec: eval - results - adaptive}
When nodes submit a demand to the network, they may either request a fixed or adaptive rate of packet generation.
One may expect that requesting an adaptive rate of packets is beneficial, as any time that the demand spends queuing is accounted for when determining how frequently PGAs need to be scheduled. 
For adaptive rates therefore, the process of queuing should not affect the probability that a session obtains minimal service, and the value of $p^{MS}$ for adaptive rate demands should be greater than that of fixed rate demands for all $\lambda$. 

In both the peer-to-peer \verb|MDA| and the client-server \verb|CKA| simulations, this expected behaviour is observed for small values of $\lambda$, with the value of $p^{MS}$ matching the value of the best fixed rate demands. 
However, once the value of $\lambda$ passes a critical value (0.001 for \verb|MDA| and $10^{-5}$ for \verb|CKA|), the value of $p^{MS}$ for the adaptive rate demands is less than that for at least one of the fixed rate demands. 

This reversal can be explained by considering the effect that queuing has on the rate at which PGAs are scheduled, $R_\texttt{attempt}$, for adaptive rate demands. 
Recall that when end nodes request an adaptive rate, the central controller will attempt to schedule at least $N_{min}$ (equal to 850 for both test applications) PGAs in the time before the demand expires.
This ensures that given a demand is accepted for scheduling, the session will obtain minimal service with probability at least $1-\epsilon_{service}$.
The consequence of this is that the longer an adaptive rate demand waits in the demand queue, the greater the resulting value of $R_{\texttt{attempt}}$ will be.
This in turn leads to each PGT contributing more PGAs to a schedule, and utilising more of the resources in the network. 
Therefore, fewer PGTs can be scheduled in any given network schedule, leading to even longer queuing times and creating a feedback loop. 
In particular, more demands will expire before being scheduled, leading to the observed decrease in the value of $p^{MS}$.
This queuing behaviour is also reflected in Figures~\ref{fig:eval-results-p2p_mda-latency} and \ref{fig:eval-results-cs_bqc-latency}, where a (sharp) increase in the value of $\bar t_{queue}$ occurs as $\lambda$ passes the critical values above.

\subsubsection{Best Fixed Rate for Nodes to Request}
\label{sssec:eval - results - fixed rate comparison}

End nodes may alternatively derive some benefits by requesting a fixed rate of packet generation.
For instance, the rate of packet generation will be independent of any other network traffic, and can be chosen to reflect other requirements of the application not captured by, say the expiry time. 
For example, suppose that Alice and Bob want to generate a QKD key within the next half hour, but once they start generating the key, they want to complete key generation within a minute. 
This desired behaviour could be captured by requesting a high fixed rate of generation, substantially larger than the minimum rate to generate the required packets across the full half hour before the application session expires. 
Furthermore, as we have seen in the previous section, there are even some network conditions where requesting a fixed rate will lead to a greater likelihood of obtaining minimal service from the network. 

From the results of the peer-to-peer \verb|MDA| simulations in Figure~\ref{fig:eval-results-MDA}, we observe that it is beneficial for end nodes to request a lower rate of packet generation. 
This not only leads to an increased value of $p^{MS}$, but also shorter queuing times and latency before being scheduled. 
In comparison, in the client-server \verb|CKA| simulations, we observe the opposite effect.
Whilst increasing the requested rate of packets does increase the queuing time, the value of $p^{MS}$ also increases (it is exactly 1 for all $\lambda$).

We can explain this behaviour as follows:
In general it is beneficial to request a higher rate of packet generation.
Requesting a higher rate of packet generation will result in more PGAs being scheduled in any given time interval, increasing the likelihood of obtaining minimal service.
In particular, this also makes them more resilient against having to wait in the demand queue. 

However, higher rates also results in longer queuing times through greater utilisation and PGA contributions to a schedule. 
Therefore, there is a critical value of $R$, depending on the demands and demand submission characteristics, where the effects from extended queuing times starts to outweigh the benefits of potentially more PGAs being scheduled.
This is precisely what we see in Figure~\ref{fig:eval-results-p2p_mda-prop_min_service} with the fixed $R=0.2$Hz demands for peer-to-peer \verb|MDA|.
On the other hand, for the client-server \verb|CKA|, $R = 0.001$Hz is still below this critical value of $R$, and so the demands gain all the benefits of scheduling PGAs at a higher rate. The existence of such a critical value of $R$ is not surprising, as networks with control of limited numbers of resources are well known to have finite capacity regions (e.g. \cite{FlowControlI, gauthier_control_2023})

\section{Conclusions and Future Work}

We have designed a novel architecture for a quantum network which allows for the integration of network scheduling with local program execution.
This is achieved by introducing an application-motivated demand format for end nodes to request packets of entanglement generation.
This is in contrast to more limited demand formats in previous work.
In such demands, one requests a rate of end-to-end entangled link generation, which fails to capture some of the requirements for executing applications. 

We also defined a scheme for producing network schedules in a format which can be used by the local execution environment on an end node to more efficiently schedule local operations.
To do this, we introduced packet generation tasks and packet generation attempts to allow the central controller of the quantum network to effectively allocate the use of shared resources in order to satisfy submitted demands. 

We presented an example implementation of our architecture using a network scheduler based on EDF scheduling. 
Using this implementation, we performed numerical simulations of our architecture on a star-shaped network.
The results of these numerical simulations highlight that the architecture successfully provides application sessions with minimal service, both for measure-directly and create-and-keep type applications. 
Furthermore, we have seen that there is a need for smart admission control, both in selecting which demands are accepted upon arrival and in selecting which queued demands to accept for scheduling. 
Finally, according to the performance metric of the proportion of sessions obtaining minimal service, we have seen that there is no single best fixed rate to request, but rather the optimal rate depends on current traffic on the network.

Our contributions open up many opportunities for future research. 
Each of the phases of the architecture, from admission control, to queuing, to computing the network schedule, requires a tailored algorithm. 
Design and optimisation of various possibilities for each type of algorithm merits in depth investigation. 
The properties of such algorithms directly impact the performance of the network. 
In particular, the choice of scheduling algorithm will have a great effect on the specific performance guarantees which the network can promise to the nodes. 

We have shown that there is a need for efficient admission control. 
An algorithm for admission control should be tailored to enforce the desired network behaviour and to account for the expected traffic on the network.
Given a target set of performance metrics, the impact of an admission control algorithm depends strongly on the scheduling algorithm.
Thus, we expect scheduling algorithm design to inform the design of admission control algorithms, perhaps following some of the same ideas as in \cite{steenhaut_scheduling_1997}.

We also showed that whilst adaptive rates may seem like a good idea, they require extra care to handle to avoid the unintended consequence of leading to overload on the network.
Whilst such requests are good at adapting to momentary variations in the load on the network, if the increased demand is sustained then the benefits are lost. 
This is consistent with previous work on rate control algorithms in both the classical and quantum domain, which suggests that for variable request rates to be effective they need to be combined with a rate control algorithm (see for example \cite{gauthier_control_2023, FlowControlI}).
It remains to determine conditions on when precisely when this occurs, and bounds on how much the value of $R_\texttt{attempt}$ otherwise may vary by without causing an overload. 

Similarly, the optimal fixed rate which nodes should request merits further investigation.
We have shown here that in some cases it is better to request a higher fixed rate, and in other cases it is better to request a lower fixed rate. 
However, it remains to determine percisely what is causing this bifurcation, and what the optimal rate to request is.

 \label{sec: conclusions and future work}

\section{Acknowledgements}
TB, HJ, SG and SW acknowledge funding from the Quantum Internet Alliance (QIA). 
QIA has received funding from the European Union’s Horizon Europe research and innovation programme under grant agreement No. 101102140.
SW also acknowledges funding from NWO VICI. We thank Wojciech Kozlowski and Ingmar te Raa for critical feedback on the content of this manuscript and thank Bart van de Vecht for advice about Qoala.

\small
\bibliographystyle{IEEEtran}
\bibliography{main}
\clearpage
\tableofcontents
\listoftheorems[ignoreall, show={define}, title=List of Definitions]
\listofalgorithms
\listoffigures
\listoftables
\appendix
\normalsize
\onecolumn

\begin{center}
{\Large\scshape \textbf{Appendices}}
\end{center}

\section{Further Discussion about Architecture}

\subsection{Discussion of time-step induced issues}
\subsubsection{Periodicity of entanglement generation attempts}
There are a few issues which may arise from the fact that entanglement generation can be a periodic process.
For instance, if the duration of a PGA is not a multiple of this period, then the node will not be able to exactly match what the network has scheduled, perhaps even leading to de-synchronisation.
In order to mitigate this, we propose that the end nodes should only do entanglement generation during times \textit{within} the allocated time periods.
In particular, this means that if they would exceed the end of a PGA with a given entanglement generation attempt, it should not be carried out unless the node is immediately starting another PGA\@.

To show that this is not detrimental to performance, we consider two scenarios differentiated by when packet generation succeeds within a PGA. 
If a packet $(w,s,F)$ is created before the final attempt, then we expect that the end nodes will immediately begin executing local operations. 
In this case, it is irrelevant at what time the final scheduled attempt to generate entanglement would have been, as a packet has already been generated. 
Suppose instead a packet has not been generated before the final scheduled attempt.
In this case, as each attempt to generate an entangled link has a very low probability to succeed, loosing a single attempt at creating a link will not have a material impact on the overall expected performance of a PGA\@.

\subsubsection{Alignment between network and nodes}
There may also be situations where the node operates on its own time-slotted schedule where the start of a PGA does not align with the start of a time slot at the node.
In this case again we expect that the node would execute the PGA within the local time slots which fall within the allocated times.

This could be mitigated by ensuring that the network scheduling algorithm schedules start times in such a way that they coincide with local time slots. 
However, this would require both that the nodes communicate such information to the central controller, as well as a level of synchronisation beyond what is reasonable above the physical level of the network.

\subsection{Performance Limitations}

The utilisation bound and the choice of scheduling interval are two potential sources of limitations on the performance of our scheduled.  
The utilisation limit is inherent; any architecture will never be able to extract more than a utilisation of 1 out of any resource, so this will always be a limitation in any network architecture.

On the other hand, the choice of the scheduling interval can impact performance in two separate ways. 
Firstly, if the deployed scheduler is too computationally complex, then in order to ensure that the schedule is distributed on time, we will need to limit how many demands or PGAs are included in a given schedule. 
This means fewer demands can be served in a given schedule, therevy reducing performance. 
This effect will become more apparent for shorter scheduling intervals, so this can be mitigated by extending the scheduling interval. 

However, extending the scheduling interval may impact the performance elsewhere. 
In particular, if the scheduling interval it too long with respect to the value of $\tmaxdur$ for a given class of demands, then the inherent latency will greatly decrease the performance of such demands. 

Therefore, the choice of the length of the scheduling interval needs to be made carefully, considering typical demands on the network and the complexity of the selected scheduler to avoid limiting the performance which can be extracted from the network architecture.

\subsection{The End-to-End Principle}
In classical networking, a good principle to adhere to when designing network architectures is the so-called \textit{end-to-end} principle, where any process required to ensure the correctness of a network application is performed by the end users and not by the network \cite{saltzer_end--end_1984}. 
In the context of a quantum network, one could interpret this as meaning that the network should not take steps to ensure that entangled links are generated sufficiently close together, nor ensure that at least some minimum number of instances of the application are executed. 
However, we would appear to be violating this by including packets in the demand, and allowing nodes to submit a request for a variable rate of packet generation.

With regards to adaptive rate requests, the network does take some responsibility for ensuring that there are enough PGAs to give the session a probability $1 - \varepsilon_{service}$ of obtaining minimal service and executing at least $\ninst$ instances. 
However, as the required rate is calculated when the demand is accepted for scheduling rather than based on what happens when the schedule is executed, there is no need for extra communication or processing to facilitate this. 

Furthermore, given that it will often take many scheduling intervals for a given session to obtain minimal service, it may be advantageous for the network to ensure that sessions obtain minimal service if they get any service.
In particular, this would avoid the scenario where a session misses minimal service by a couple of instances.
Such a scenario results in a lot of wasted time in both the network schedule and node schedules because the demand will require re-submission and one or more previous PGAs may have used time in both network and node schedules that could have been otherwise allocated to other demands.

To justify the inclusion of packets of entanglement in the demand, recall that we wish to compute network schedules in an \textit{a priori} manner, in order to allow the node schedulers to take them as an input. 
In order to create efficient and effective schedules, however, the network does need to know about what packet of entanglement is being created, in particular to ensure that any scheduled PGAs have a sufficient duration. 
Furthermore, as the central controller doesn't get any information from the nodes/network as the schedule is executed, the network does not undertake any additional checks, e.g. determining whether or not a packet was actually generated.
This means that including packets in the demand results in an increase in network performance without encumbering the central controller with additional tasks. 
Additionally, the responsibility still lies with the nodes to adhere to the packet form they requested, rather than being enforced by the central controller.

Therefore, by permitting packets and adaptive rates of packet generation. the additional responsibilities of the network are minimal, and each has the potential to increase the overall performance of the network.

\clearpage
\section{Notation}
    Notation used throughout this paper. Entries which fall into multiple categories are only shown once.
    
    \begin{tabular}{c|l}
        \textit{\textbf{Symbol}} & \textit{\textbf{Definition}} \\\hline
        \multicolumn{2}{c}{\textbf{Application Sessions}}\\\hline
         $\mathcal{S}$ & Application session\\
         $\mathcal N$ & Set of nodes in the network\\
         $\texttt{App}$ & placeholder for a quantum application\\
         $\ninst$ & Minimum number of application instances required for minimal service\\
         $\texpiry$ & Expiry time of an application session/demand/packet generation task\\\hline

        \multicolumn{2}{c}{\textbf{Packets of Entanglement}}\\\hline
            
         $p$ & packet of entanglement\\
         $w$ & time window of packet\\
         $s$ & number of pairs in a packet\\
         $F$ & (average) minimum fidelity of links generated as part of a packet\\\hline

        \multicolumn{2}{c}{\textbf{Demands}}\\\hline
        
         $\mathcal{D}$ & Network Demand\\
         $R$ & Requested rate of packet generation\\
         $\tminsep$ & Minimum time between two packet generation attempts\\\hline

        \multicolumn{2}{c}{\textbf{Packet Generation Tasks}}\\\hline
         $\tau$ & Packet generation task (PGT)\\
         $\tau_{i,j}$ & The $j$th packet generation attempt (PGA) stemming from PGT $\tau$\\
         $E$ & Execution time of a PGA\\
         $R^\texttt{attempt}$ & (Minimum) rate at which PGAs should be scheduled\\
         $\rho$ & Resources required to execute a PGA.\\\hline

         \multicolumn{2}{c}{\textbf{Network Schedules}}\\\hline
         
         $T_{SI}$ & Duration of the scheduling interval \\\hline
         
         \multicolumn{2}{c}{\textbf{Evaluation}}\\\hline
         
         $\epsilon_{service}$ & Probability that a session does \textbf{not} obtain minimal service\\
         $p_\texttt{packet}$ & Probability that a packet is generated in a given PGA\\
         $p_{gen}$ & Probability that an attempt to create a single entangled link succeeds\\
         $t_\texttt{submit}$ & Time at which a demand is received by the central controller\\
         $\tmaxdur$ & Maximum duration of an application session. Used for dynamically setting expiry times.\\\hline

         \multicolumn{2}{c}{\textbf{Applications}}\\\hline

         \verb|MDA| & Measure Directly Application, based on QKD\\
         \verb|CKA| & Create and Keep Application, based on BQC\\\hline

         \multicolumn{2}{c}{\textbf{EDF Scheduler}}\\\hline

         $T$ & (Effective) period of a PGT\\
         $\sigma$ & Offset from $t=0$ of a PGT\\
         $r_{i,j}$ & release time of PGA $j$ from PGT $i$\\
         $d_{i,j}$ & deadline of PGA $j$ from PGT $i$\\
         $s_{i,j}$ & start time of PGA $j$ from PGT $i$\\
         $c_{i,j}$ & completion time of PGA $j$ from PGT $i$\\
         $t^*$ & scheduler decision time\\\hline

         \multicolumn{2}{c}{\textbf{Admission Control}}\\\hline
         
         $U_\tau$ & Utilisation of PGT $\tau$\\
         $\mathcal{U}_r$ & utilisation of resource $r$\\
         $\hat U$ & Utilisation bound\\
         $\alpha_C$ & Proportion of the scheduling interval allocated to compute the network schedule within.\\ 
         $\mathcal T$ & Set of PGTs from which PGAs should be scheduled in the next network schedule.

    \end{tabular}
 \label{app: notation}

\section{Who knows what}

\subsection{Central Controller}
{\centering
\begin{tabular}{|c|c|}
\hline
    Knows & Doesn't know \\\hline
 Probability of PGA failure $\epsilon_{packet} = 1-p_{packet}$ & If a PGA is/was successful\\
 Network topology & What application(s) are being run\\
Network traffic & \\
 EGPs available on the network & \\
Capabilities of internal components e.g. repeater chains. & \\\hline
\end{tabular}

\subsection{Nodes}
\begin{tabular}{|c|c|}
\hline
    Knows & Doesn't know \\\hline
Local hardware capabilities &  Full network topology\\
Local utilisation & Network traffic\\
Application program(s) [Number of pairs required, local gates] & EGPs on network\\
If an instance executes (successfully) & $\varepsilon_{packet}$ (up to empirical deductions)\\
Identity of nearest neighbour \& metro hub & Application sessions being run on other nodes\\

\hline
\end{tabular}}

 \label{app: who knows what}

\section{Code}
\label{app: code}

\subsection{Figures}\label{subsec:code guide - figures}
\begin{table}[!h]
    \begin{tabularx}{\textwidth}{P{3cm}|C|C|C|}
        Figure Number & Dataset & UUID & Script\\\hline
        \ref{fig:eval-results-MDA} & 20240709-1100 & 01379bb0-f27b-4e88-8ee1-cfb739703615 & \lstinline|combined_plotting.py| \\\hline
        \ref{fig:eval-results-BQC} & 20240711-100 & 6d415ef2-f092-440c-850d-2d1c50f08640 & \lstinline|combined_plotting.py| \\\hline
    \end{tabularx}\label{tab:figures table}
\end{table}

\section{Methods for creating Packet Generation Tasks}
\subsection{Approximations from Naus '82 for determining the length of PGAs}

\subsubsection{Motivation}\label{app: Naus Approximation}

When determining how long a PGA should be, we need to be able to calculate the probability of a packet of entanglement being generated in some timeframe consisting of a known number of trials. 
This is precisely the sort of problem which is addressed in the field of \textit{scan statistics}, which is dedicated to looking at the probability that random events are grouped together \cite{glaz_scan_2001}.
In particular, the process of generating a packet is equivalent to the \textit{generalised birthday problem}, which looks at the probability of getting $k$ events (or successes) in a window of size $m$. 
Exact formulae as well as approximations exist to calculate these probabilities, we use the one in the following section which is due to Naus \cite{naus_approximations_1982}.

\subsubsection{result}

The following theorem and discussion is due to Naus \cite{naus_approximations_1982}:

\begin{theorem}
Let $T_{k,m}$ be the time at which we first observe $k$ events in a window of size $m$.
Then if we write \begin{align*}
    P'(k | m,N,p) &= \mathbb P[T_{k,m} < N]\\
                  &= 1 - Q'(k| m; N ; p),
\end{align*} and abbreviate $Q'(k| m; Lm; p)$ as $Q_L'$, then \begin{equation}
    Q_L' \approx Q_2'\left(\frac{Q_3'}{Q_2'}\right)^{\frac{N}{m}-2}\label{eq: Naus Approximation for PGA success probability}
\end{equation}
\end{theorem}

Naus also gives formulae for calculating $Q_2'$ and $Q_3'$ exactly:

Let \begin{align*}
    b(k;m,p) &= \begin{pmatrix}
        m \\ k
    \end{pmatrix} p^k(1-p)^{m-k}, ~ & 
    F_b(r;s,p) &= \left\lbrace\begin{matrix}
        \sum_{i=0}^r b(i;s;p) & r = 0,1,...,s \\
        0 & r < 0
    \end{matrix}\right..
\end{align*} 
 Then for $2<k<N$, $0<p<1$, we have \begin{equation}
    Q_2' = \left(F_p(k-1;m,p)\right)^2 - (k-1)b(k;m,p)F_b(k-2;m,p) + mpb(k;m,p)F_b(k-3;m-1,p)
\end{equation} and \begin{equation}
    Q_3' = \left(F_b(k-1,m,p)\right)^3 - A_1 + A_2 + A_3 - A_4
\end{equation} where \begin{align*}
    A_1 &= 2b(k;m,p)F_b(k-1;m,p) \bigg\lbrace(k-1)F_b(k-2;m,p) - mpF_b(k-3;m-1,p)\bigg\rbrace\\
    A_2 &= \frac{1}{2}b_k^2\left( (k-1)(k-2)F_b(k-3; m,p) - 2(k-2)mpF_b(k-4;m-1,p) + m(m-1)p^{2}F_b(k-5; m-2,p) \right)\\
    A_3 &= \sum_{r=1}^{k-1}b_{2k-r}F_b^2(r-1; m, p)\\
    A_4 &= \sum_{r=2}^{k-1}b_{2k-r}b_r((r-1)F_b(r-2; m,p) - mpF_b(r-3; m-1, p))
\end{align*}

\subsubsection{Impact}
Given this approximation, we can calculate the probability that a packet is produced in a PGA of length $N$ timesteps. 
From here, we can use a method such as interval bisection to calculate the length of PGA required to exceed the desired value of $p_\texttt{packet}$.

One implication of using this method for calculating the length of PGAs is the scaling with regards to different parameters.
In particular, if we tighten the window or increase the value of $p_\texttt{packet}$, then the required length of the PGA grows rapidly, much faster than the value of $R_{\texttt{attempt}}$ decreases. 

\subsection{Calculating Minimum Rate using Hoeffding's inequality.}
\label{app: PGT creation - Min Rate Hoeffding}
Let packet generation attempts occur at rate $R$, and succeed with probability $p$. Let the session have an acceptable probability of failure of at most $\epsilon_{service}\ll 1$, an expiry time of $t_\texttt{expiry}$ and require $\ninst$ instances to be successfully executed to obtain minimal service. Let $N = R\times\texpiry$, and let $\ninst = \alpha N$. Let $S_N$ be the number of successfully executed instances.  

Let $(X_i)_{i=1}^N\overset{iid}{\sim}\text{Bernoulli}(p)$. Then $S_N\overset{d}{=}\sum_{i=1}^NX_i.$ Then \begin{align}
    \mathbb P[\text{not minimal service}] &= \mathbb P[S_N < \ninst]\\
    &= \mathbb P[\mathbb E[S_N] - S_N > \mathbb E[S_N] - \ninst]\\
    &= \mathbb P[pN - S_N > N(p-\alpha)]\\
    &\leq \exp(-2N(p-\alpha)^2)\label{eq: min rates Hoeffding}
\end{align} where the last inequality is by Hoeffding's inequality \cite{hoeffding_probability_1963}.

The minimum number of packet generation attempts required is then given by \begin{equation}
\min\left\lbrace N~|~\epsilon_{service} > \exp(-2N(p-\alpha)^2) \wedge N > \ninst\right\rbrace
\end{equation}
from which we can then recover the minimum rate.

\clearpage
\section{Additional Evaluation details}
\label{app:Additional Eval Details}
\subsection{Entanglement Generation Model}
\subsubsection{Werner States}
We assume that all generated links are Werner states~\cite{werner_quantum_1989}, and can be written in the format \begin{equation}
    \rho = \frac{1-F}{3}\mathbb{I}_2 + \frac{4F-1}{3}|\psi\rangle\langle\psi|.
\end{equation}
where 
\begin{equation*}
    |\psi\rangle = \frac{|00\rangle + |11\rangle}{2}
\end{equation*}
and $F$, the fidelity of $\rho$, is fixed.

The precise values we use are given in Table~\ref{tab:network paramters main}.

\subsubsection{Entanglement Generation}
For our model, we assume that there are network-wide timeslots, at the end of which an entangled link between the outer nodes and the central junction node is generated with known probability at a given fidelity. 
These are then instantaneously and deterministically swapped, if possible, to create the desired end-to-end entangled links. 
We also assume that the central node has no memories, and so it cannot store links for longer than one time-slot.

This means that there is only one possible scheme for generating end-to-end links, which is used by all pairs of nodes. 

We also assume that all links are homogeneous, in particular that they all produce links of the same average fidelity. 

\subsection{Choice of cap on number of PGAs per schedule}
\label{app: eval supp - PGA cap}
\begin{figure}
    \centering
    \includegraphics{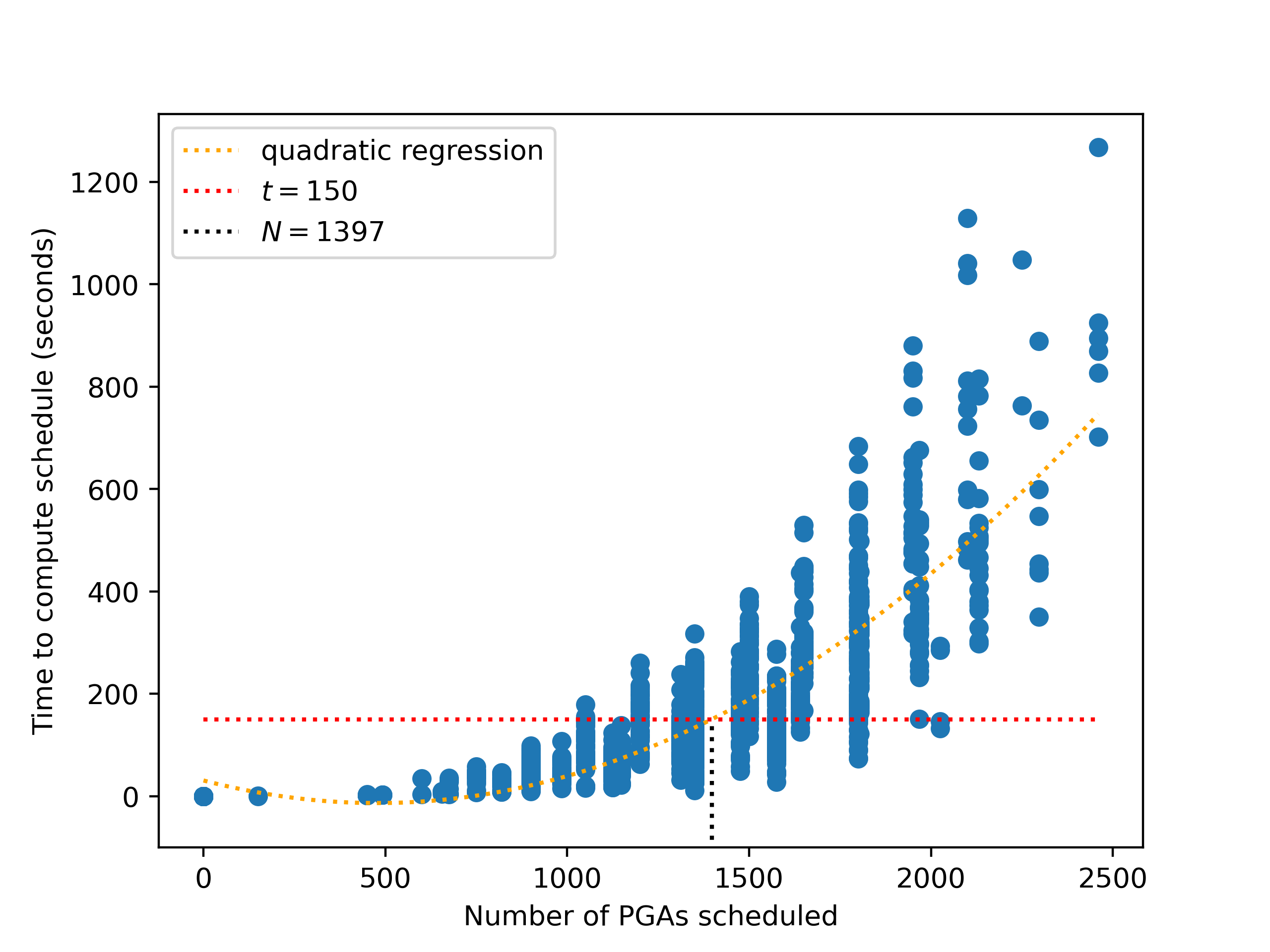}
    \caption[Times taken for our computational server to compute the network schedule.]{Times taken for our computational server to compute network schedules. The server has an Intel\textsuperscript{\textregistered} Xeon\textsuperscript{\textregistered} Gold 6230 CPU, with each core running at a maximum clock speed of 3.9GHz, and 189GB of random access memory. Each simulation was run on a single core, with up to 40 simulations being run in parallel. The regression modelling was performed using the \texttt{numpy} Python module}
    \label{fig:schedule computation times}
\end{figure}

To be able to estimate a suitable choice of cap on the number of PGAs per schedule, we computed {\red 1125} network schedules. 
This was achieved by running the same simulation used for the evaluation, without a cap on the number of PGAs per schedule, and recording the number of PGAs which were scheduled and the time each schedule took to compute.
We used the QKD application with requested packet rates of $R\in\{1.0,1.1,1.5\}$ and renewal rates $\lambda\in\{0.001, 0.0015, 0.002\}$. 
Each simulation lasted 7800 simulated seconds, for a total of 25 schedules computed in each simulation. We repeated each simulation 5 times. 

Our implementation of the scheduling algorithm has complexity $O(N^2)$ where $N$ is the number of PGAs to be scheduled. 
We therefore fitted a quadratic curve to the data and used this to obtain an estimate of the maximum number of PGAs which could be scheduled in under 150s (half the scheduling period).
This can be seen in Figure~\ref{fig:schedule computation times}.

From this we obtained an estimate of 1397, which we then rounded up to 1500 for neatness. 
This was done in part as we still had some uncertainty about the performance of the server whilst performing these simulations.
In particular on other hardware we observed lower computational times, in which case the cap would be greater. 
Furthermore, in the \verb|MDA| simulations, each PGT typically contributes between 150 and 300 PGAs to a schedule, and so increasing the cap by less than 150 will not allow more demands to be accepted from the queue.

\subsection{Choice of Expiry times and Rates}
\label{app:eval:expiry-rates}
Using the method in \ref{app: PGT creation - Min Rate Hoeffding}, we find that to have $\epsilon_{service} = 10^{-5}$, 850 PGAs should be scheduled. 
Therefore, we choose the rates to simulate such that at least 850 PGAs can be scheduled within $t_\texttt{max duration} - T_{SI}$ where $T_{SI}$ is the scheduling interval.

In particular, we choose to simulate the end nodes requesting: \begin{itemize}
    \item An adaptive rate ($R=0$).
    \item Fixed rates of: \begin{itemize}
        \item Approximately the minimum rate required for 850 PGAs to be scheduled in $\tmaxdur$, $R^{min}_{\epsilon_{service}}$.
        \item approximately $2R^{min}_{\epsilon_{service}}$
    \end{itemize}
\end{itemize} 
We choose to look at demands requesting a fixed packet generation rate of $R_{\epsilon_{service}}^{min}$, as it serves as a good comparison to the adaptive rate (which would use this rate in the absence of queuing).
The choice of $2R_{\epsilon_{service}}^{min}$ gives us a motivated choice of a higher rate which we can compare against to see the impact of the choice of $R$ on the performance metrics. 

To choose the values of $\tmaxdur$, we look to the utilisation of the resulting PGTs when requesting the lowest fixed rate, $R_{\epsilon_{service}}^{min}$. 
As there are only 6 nodes in the network, if we choose a value of $\tmaxdur$ sufficiently long that the minimum rate gives a utilisation less than $\resourceutilisationbound/5$, then all demands will always be immediately accepted and everyone will get minimal service almost surely.
Likewise, if we set the value of $\tmaxdur$ too short, then only a couple of demands will be serviceable at a time without violating either \ref{sac rule:utilization} or \ref{sac rule:computationBound}, due to the high utilisation required. 
In a given deployment, we would expect that this is not desirable behaviour, and so such demands would be predominantly filtered out by the demand registration. 
Therefore, we choose the values of $\tmaxdur$ such that the utilisation of tasks requesting the slowest fixed rate is approximately 0.2, as then almost all demands can be satisfied simultaneously whilst the demand queue can still exert some influence over the performance of the network. 
We therefore choose $\tmaxdur^{\texttt{MDA}} = 2100s$ and $\tmaxdur^{\texttt{CKA}} = 4$ days. 
Note that with a scheduling interval $T_{SI}^{\texttt{MDA}} = 300$s, the choice of $\tmaxdur^{\texttt{MDA}} = 2100s$ results in an effective schedulable time of half an hour. 
For both \verb|MDA| and \verb|CKA|, we see the same trends for different values of $\tmaxdur$ whilst the same pressures from $\resourceutilisationbound$ and the PGA cap exist. 
 
\subsection{Scheduler Model}
We use the network scheduler as described in \ref{subsec: implimentation - Network Scheduling}, with a scheduling interval of 300 seconds for the \verb|MDA| simulations and 3600 seconds (1 hour) for the \verb|CKA| simulations. 
We take a longer scheduling interval for the \verb|CKA| simulations as these demands last for much longer to serve than the \verb|MDA| demands.
Subsequently, this means that sessions are renewed on a much longer timescale and so the network state (schedule and demand queue) changes on a longer time scale. 

\subsection{Qoala files}
\label{app: eval - qoala files}
\subsubsection{Measure Directly Application (\texttt{MDA})}

\lstdefinestyle{mystyle}{
    captionpos=b,
    numbers=left,
    basicstyle=\ttfamily\footnotesize,
    breaklines=true,
    frame=line,
}
\lstset{style=mystyle}
\begin{lstlisting}[caption=Qoala file for Alice]
META_START
    name: alice
    parameters: bob_id
    csockets: 0 -> bob
    epr_sockets: 0 -> bob
META_END

^b1 {type = QC}:
    tuple<m0> = run_request(tuple<>) : req

^b2 {type = CL}:
    return_result(m0)


REQUEST req
  callback_type: wait_all
  callback:
  return_vars: m0
  remote_id: {bob_id}
  epr_socket_id: 0
  num_pairs: 1
  virt_ids: all 0
  timeout: 1000
  fidelity: 1.0
  typ: measure_directly
  role: create
\end{lstlisting}
\begin{lstlisting}[caption=Qoala file for Bob]
META_START
    name: alice
    parameters: bob_id
    csockets: 0 -> bob
    epr_sockets: 0 -> bob
META_END

^b1 {type = QC}:
    tuple<m0> = run_request(tuple<>) : req

^b2 {type = CL}:
    return_result(m0)


REQUEST req
  callback_type: wait_all
  callback:
  return_vars: m0
  remote_id: {bob_id}
  epr_socket_id: 0
  num_pairs: 1
  virt_ids: all 0
  timeout: 1000
  fidelity: 1.0
  typ: measure_directly
  role: create
\end{lstlisting}
\subsubsection{Create and Keep Application (\texttt{CKA})}
\begin{lstlisting}[caption=Qoala file for Client]
META_START
    name: client
    parameters: server_id, alpha, beta, theta1, theta2
    csockets: 0 -> server
    epr_sockets: 0 -> server
META_END

^b0 {type = CL}:
    csocket = assign_cval() : 0

^b1 {type = QC}:
    run_request(tuple<>) : req0
^b2 {type = QL}:
    tuple<p2> = run_subroutine(tuple<theta2>) : post_epr_0
^b3 {type = QL}:
    tuple<p1> = run_subroutine(tuple<theta1>) : post_epr_1

^b4 {type = CL}:
    x = mult_const(p1) : 16
    minus_theta1 = mult_const(theta1) : -1
    delta1 = add_cval_c(minus_theta1, x)
    delta1 = add_cval_c(delta1, alpha)
    send_cmsg(csocket, delta1)

^b5 {type = CC}:
    m1 = recv_cmsg(csocket)

^b6 {type = CL}:
    y = mult_const(p2) : 16
    minus_theta2 = mult_const(theta2) : -1
    new_beta = bcond_mult_const(beta, m1) : -1
    delta2 = add_cval_c(new_beta, minus_theta2)
    delta2 = add_cval_c(delta2, y)
    send_cmsg(csocket, delta2)

    return_result(p1)
    return_result(p2)


SUBROUTINE post_epr_0
    params: theta2
    returns: p2
    uses: 0
    keeps:
    request:
  NETQASM_START
    load C0 @input[0]
    set Q0 0
    rot_z Q0 C0 4
    h Q0
    meas Q0 M0
    store M0 @output[0]
  NETQASM_END

SUBROUTINE post_epr_1
    params: theta1
    returns: p1
    uses: 1
    keeps:
    request:
  NETQASM_START
    load C0 @input[0]
    set Q0 1
    rot_z Q0 C0 4
    h Q0
    meas Q0 M1
    store M1 @output[0]
  NETQASM_END

REQUEST req0
  callback_type: wait_all
  callback:
  return_vars:
  remote_id: {server_id}
  epr_socket_id: 0
  num_pairs: 2
  window: 5_000_000
  virt_ids: increment 0
  timeout: 1000
  fidelity: 1.0
  typ: create_keep
  role: create
\end{lstlisting}
\begin{lstlisting}[caption=Qoala file for Server]
META_START
    name: server
    parameters: client_id
    csockets: 0 -> client
    epr_sockets: 0 -> client
META_END

^b0 {type = CL}:
    csocket = assign_cval() : 0
    iterations = assign_cval() : 0

^b1 {type = QC}:
    run_request(tuple<>) : req0
^b2 {type = QL}:
    run_subroutine(tuple<>) : local_cphase

^b3 {type = CC}:
    delta1 = recv_cmsg(csocket)

^b4 {type = QL}:
    tuple<m1> = run_subroutine(tuple<delta1>) : meas_qubit_1

^b5 {type = CL}:
    send_cmsg(csocket, m1)

^b6 {type = CC}:
    delta2 = recv_cmsg(csocket)

^b7 {type = QL}:
    tuple<m2> = run_subroutine(tuple<delta2>) : meas_qubit_0

^b8 {type = CL}:
    return_result(m1)
    return_result(m2)



SUBROUTINE local_cphase
    params: 
    returns: 
    uses: 0, 1
    keeps: 0, 1
    request: 
  NETQASM_START
    set Q0 1
    set Q1 0
    cphase Q0 Q1
  NETQASM_END

SUBROUTINE meas_qubit_1
    params: delta1
    returns: m1
    uses: 0, 1
    keeps: 0
    request: 
  NETQASM_START
    load C0 @input[0]
    set Q1 1
    rot_z Q1 C0 4
    h Q1
    meas Q1 M0
    store M0 @output[0]
  NETQASM_END

SUBROUTINE meas_qubit_0
    params: delta2
    returns: m2
    uses: 0, 1
    keeps:
    request: 
  NETQASM_START
    load C0 @input[0]
    set Q0 0
    rot_z Q0 C0 4
    h Q0
    meas Q0 M0
    store M0 @output[0]
  NETQASM_END

REQUEST req0
  callback_type: wait_all
  callback: 
  return_vars: 
  remote_id: {client_id}
  epr_socket_id: 0
  num_pairs: 2
  window: 5_000_000
  virt_ids: increment 0
  timeout: 1000
  fidelity: 1.0
  typ: create_keep
  role: receive
\end{lstlisting}

\subsection{Sensitivity to parameters}
\subsubsection{Dummy Network Scheduler}
In order to speed up our simulations when testing the sensitivity of our simulations to the chosen parameters, we do not explicitly calculate the network schedule each time. 
To determine when a session obtains minimal service, it is simply required to know how many PGAs for a particular demand were scheduled in a given network schedule/scheduling interval. 
From this we can obtain the number of packets which were actually generated by sampling a $\mathrm{Binomial}(N,p_\texttt{packet})$ distribution, and thereby establish whether the session obtained minimal service in that scheduling interval. 

As our network scheduler schedules precisely one PGA per period of the PGT, given which demands/PGTs have been admitted by the scheduler admission control, we can calculate how many PGAs will be scheduled without having to calculate the schedule directly. 
Then as described above we can establish how many packets were generated and thus whether a session obtained minimal service. 

We use this method of simulating the network scheduling for the additional data gathered for sensitivity testing of the simulation parameters. We also validated it against computing the network schedule and saw a perfect match for the same datasets.

\subsection{Calculating the expected queuing time}
\label{app: eval - expected queuing time}

We use the notation from \S~\ref{ssec: eval - session model}. Let $t_{renew}\sim\mathrm{Exponential}(\lambda)$.
Then the time which a demand $\mathcal{D}$ would sit in the demand queue for in the absence of other demands is $Q \equiv T_{SI} - t_{renew} \mod T_{SI}$. 
For convenience we write $T_{SI} = T$, and let $t_\mathcal D$ be the time demand $\mathcal{D}$ is submitted. 
We then calculate:
\begin{equation}  
    F_Q(\tau) = \mathbb P[Q\leq \tau] = \mathbb P\left[t_\mathcal{D}\in\bigcup_{k=1}^\infty[kT-\tau, kT]\right] = \sum_{k=1}^\infty\int_{kT-\tau}^{kT}\lambda e^{-\lambda x}\mathrm{d}x = \sum_{k=1}^\infty e^{-k T\lambda}(-1+e^{\lambda \tau}) = \frac{-1 + e^{\lambda\tau}}{-1+e^{\lambda T}}
\end{equation}
\begin{equation}
    \mathbb E[Q] = \int_{0}^T\tau~\mathrm{d}F_Q = \int_0^T\frac{\tau\lambda e^{\lambda \tau}}{-1 + e^{\lambda T}}~\mathrm{d}\tau = T\left( 1 - \frac{1}{\lambda T} + \frac{1    }{e^{\lambda T}-1}\right)
\end{equation}

\end{document}